\documentclass[twocolumn,prc,showpacs,superscriptaddress,preprintnumbers,nofootinbib]{revtex4}
\usepackage{amsmath,amssymb,longtable}
\usepackage{graphicx}
\usepackage{epstopdf}
\usepackage{subfigure}
\usepackage{epsfig}
\usepackage{hyperref}
\usepackage{array}
\usepackage{hyperref}
\usepackage{color,xcolor}
\usepackage{bm}
\usepackage{ulem}

\usepackage{multirow}

\begin{document}

\title{Hadron productions and jet substructures associated with $Z^0/\gamma$ \\in Pb+Pb collisions at the LHC}

\date{\today  \hspace{1ex}}

\author{Shan-Liang Zhang}
\email[]{zhangshanl@m.scnu.edu.cn}
\affiliation{Key Laboratory of Atomic and Subatomic Structure and Quantum Control (MOE), Guangdong Basic Research Center of Excellence for Structure and Fundamental Interactions of Matter, Institute of Quantum Matter, South China Normal University, Guangzhou 510006, China.}
\affiliation{Guangdong-Hong Kong Joint Laboratory of Quantum Matter, Guangdong Provincial Key Laboratory of Nuclear Science, Southern Nuclear Science Computing Center, South China Normal University, Guangzhou 510006, China.}

\author{Hongxi Xing}
\email[]{hxing@m.scnu.edu.cn}
\affiliation{Key Laboratory of Atomic and Subatomic Structure and Quantum Control (MOE), Guangdong Basic Research Center of Excellence for Structure and Fundamental Interactions of Matter, Institute of Quantum Matter, South China Normal University, Guangzhou 510006, China.}
\affiliation{Guangdong-Hong Kong Joint Laboratory of Quantum Matter, Guangdong Provincial Key Laboratory of Nuclear Science, Southern Nuclear Science Computing Center, South China Normal University, Guangzhou 510006, China.}
\affiliation{Guangdong-Hong Kong Joint Laboratory of Quantum Matter, Southern Nuclear Science Computing Center, South China Normal University, Guangzhou 510006, China}

\author{Ben-Wei Zhang}\email[]{bwzhang@mail.ccnu.edu.cn}
\affiliation{Institute of Particle Physics and Key Laboratory of Quarks and Lepton Physics (MOE), Central China Normal University, Wuhan 430079, China}

\begin{abstract}
We carry out a detailed study of medium modifications on $Z^0$/$\gamma$+hadron correlations as well as jet substructures in association with  $Z^0/\gamma$ in Pb+Pb collisions with $\sqrt{s_{NN}}=5.02$~TeV at the LHC. We utilize the Linear Boltzmann transport (LBT) model to simulate the jet-medium interactions and medium response, and an extended cluster hadronization model to investigate the non-perturbative transition of quarks and gluons into final hadrons in heavy-ion collisions. Including hadronization effect, we can well describe $Z^0/\gamma$+hadrons correlations and $Z^0/\gamma$-tagged jet substructures  in both  p+p  and Pb+Pb collisions simultaneously.
Medium modification on jet profile and jet fragmentation functions indicate that
particles carrying a large fraction of the jet momentum are generally closely aligned with the jet axis, whereas low-momentum particles are observed to have a much broader angular distribution relative to jet axis in Pb+Pb collisions due to jet-medium interactions. In particular, we find that $Z^0/\gamma$-tagged hadron correlations are sensitive to the soft particles from the dense medium  and medium response,  while jet-substructures show weak dependence on those soft hadrons with only a fraction of them falling inside the jet area.\\

\end{abstract}


\maketitle

\section{Introduction}

A novel state of the QCD matter with deconfined quarks and gluons, the quark-gluon plasma (QGP), is believed
to be produced on a fleeting timescale, $\tau \sim$1 fm/c, in high energy nucleus-nucleus collisions at the Relativistic Heavy-ion Collider (RHIC) and the Large Hadron Collider (LHC) . These quarks and gluons, known as partons, created prior to the formation of the QGP, traverse the hot-dense medium and experience elastic and inelastic scatterings with the constituents of the medium. The interactions between the hard jets and the soft medium usually reduce the energy of the incoming jet partons, which leads to the suppression of final observed hadrons and full jet yield at high transverse momentum relative to elementary proton-proton (p+p) collisions at the same colliding energy.
This process, called jet quenching, has been long regarded as one of the best hard probes to study
the properties of QGP~\cite{Wang:1991xy,Gyulassy:2003mc,Qin:2015srf,Vitev:2008rz,Vitev:2009rd,Qin:2010mn,CasalderreySolana:2010eh,Young:2011qx,He:2011pd,ColemanSmith:2012vr,Zapp:2012ak,Ma:2013pha,Senzel:2013dta,
Casalderrey-Solana:2014bpa,Milhano:2015mng,Chang:2016gjp,Majumder:2014gda,Chen:2016cof,Ru:2014yma,Dai:2018mhw,Chen:2019gqo,Zhang:2021sua,Chen:2020pfa,Liu:2021nyg,Dong:2022cuw,Wang:2022yrp,Qiu:2020xum,Rapp,Wang:2021xpv,Zhang:2021iqp,Zhang:2021xib}

Early studies of jet quenching focused on high momentum hadrons and there have been tremendous theoretical/phenomenological studies and experimental measurements on the suppression of high momentum hadrons~\cite{Adams:2003kv,Adler:2003qi,Back:2004bq,Aamodt:2010jd,CMS:2012aa,Zhang:2022rby}. In recent years at the RHIC and especially with the increasing collision energies at the LHC, productions of full jet, reconstructed from collimated clusters  in the final-state particles within a given radius defined as  $\Delta R=\sqrt{(\Delta \phi)^2+(\Delta \eta)^2}$,  such as dijet $p_\text{T}$
imbalance\cite{ATLAS:2010isq,CMS:2011iwn,STAR:2016dfv}, modifications of the jet yield in the medium~\cite{Vitev:2008rz,Adam:2015ewa,Khachatryan:2016jfl,Aad:2014bxa,Takacs:2021bpv}, electroweak boson-jet correlations~\cite{Dai:2012am,Wang:2013cia,Chen:2018fqu,Sirunyan:2017qhf,Sirunyan:2017jic,Neufeld:2010fj,Neufeld:2012df,Casalderrey-Solana:2015vaa,KunnawalkamElayavalli:2016ttl,Kang:2017xnc,Zhang:2018urd,Zhang:2021oki},  missing $p_\text{T}$ in dijet systems\cite{CMS:2015hkr}, have been intensively explored to extract the properties of QGP.
Though medium modifications on high energy inclusive hadrons are mainly  attributed to the interaction of the leading partons with the medium constituents, full jets are modified by the interactions of all of the daughter partons of a given  parton with the hot-dense medium, and also by induced medium response, the result of the excitation of the thermal partons from jet-medium interaction~\cite{Cao:2020wlm,Chen:2017zte}. Since full jets are composite objects with complex internal structures~\cite{Zhang:2021sua,Caucal:2021cfb}, jet quenching may leave prints on jet substructures as well as jet yields.

Inclusive jet shape and jet fragmentation functions had been measured by ATLAS and CMS at the LHC~\cite{Chatrchyan:2013kwa,Chatrchyan:2014ava,Aad:2014wha,Chatrchyan:2012gw}. Later, $\gamma$ triggered jet shape~\cite{CMS:2018jco} and fragmentation function~\cite{Sirunyan:2018qec,Chen:2020tbl} have also been measured and studied, which show different modification patterns compared to that of inclusive jet~\cite{Chang:2019sae,Aaboud:2019oac}.  The productions of charged particles in Pb+Pb collisions tagged with $Z^0$ boson at $\sqrt{s}=5.02 $ TeV \cite{ATLAS:2020wmg,CMS:2021otx,Yang:2021qtl}  provide additional information on the jet-medium interaction. It has been noticed that these jet observables, related to the particle number density and hadron components, are rather sensitive to hadronization processes and underlying background contamination of soft particles~\cite{Duan:2022bew,Kang:2019ahe,Aaij:2019ctd}. In addition, nuclear modification factors of charged jet recently measured by ALICE~\cite{alice:jetcone}, $\gamma$-tagged charged jet at STAR\cite{star:jetcone}, and inclusive jet at CMS~\cite{CMS:2021vui} and at ATLAS~\cite{ATLAS:2012tjt}, show different jet radius $R$ dependence.  This very interesting contrast places  a challenge to theoretical models of jet quenching, and also attaches great importance to hadronization process in heavy-ion collisions, with the quite distinct features between charged jets  and full jets in the measurements.

The above observations motivate us to develop an improved model of  non-perturbative hadronization in high-energy
nucleus-nucleus (A+A) collisions, to interface with parton-jet propagation in hot QCD medium.
The non-perturbative hadronization process in p+p reactions are usually studied via effective mechanism, such as Lund string fragmentation mechanism~\cite{Andersson:1983ia} by PYTHIA~\cite{Sjostrand:2014zea}  and cluster model~\cite{Gottschalk:1982yt,Gottschalk:1983fm,Gottschalk:1986bv,Webber:1983if}  by SHERPA and HERWIG event generators.
In A+A collisions, it has been shown that hadronizations with recombination play an important role in describing particle ratios and correlations at the
low $p_\text{T}$ in relativistic heavy-ion collisions (HIC) ~\cite{Fries:2003vb, Greco:2003xt,Hwa:2004ng,Shao:2004cn}.
 Recently Colored Hadronization~\cite{Putschke:2019yrg,JETSCAPE:2019udz}, Colorless Hadronization~\cite{Putschke:2019yrg,JETSCAPE:2019udz} have been developed to study
particle productions at large transverse momentum. A hybrid hadronization~\cite{Kordella:2020nwi,Zhao:2020wcd} is introduced as a
combination of Lund string fragmentation and recombination. To investigate heavy-flavor hadron productions
in heavy-ion collisions a model with color reconnections~\cite{Beraudo:2022dpz} has been proposed with cluster mechanism, in which partons are no longer connected by strings according to their color
charge in the final stage, in order to minimize the invariant mass of the strings.


In this paper, we generalize the cluster hadronization mechanism ~\cite{Gottschalk:1982yt,Gottschalk:1983fm,Gottschalk:1986bv,Webber:1983if}  to non-perturbative particle productions in heavy-ion collisions and develop an extended cluster model of hadronization, 
which is then utilized to study medium modification on  $Z^0$-hadron correlations and jet substructures. We simulate the $Z^0/\gamma$-tagged parton-jet events in  p+p  collisions with a MC event generator SHERPA2.24~\cite{Gleisberg:2008ta}, and these events will be embedded into the Linear Boltzmann Transport (LBT) model~\cite{Li:2010ts,He:2015pra, Cao:2016gvr} to take into account jet quenching effect in the QGP.  
In the numerical calculations, we keep track  of the whole evolution history  for elastic and inelastic scattering  during the jet propagation in hot dense medium step by step based on LBT model. After leaving the fireball the escaping partons will hadronize into baryons and mesons via the extend cluster hadronization model, and we then study in detail $Z^0/\gamma$+hadron correlations as well as $Z^0/\gamma$ tagged jet substructures, such as  jet shape, hadron number density,  and jet fragmentation function.

The outline of this paper is  as follows. In Sec.~\ref{sec:framework} , we introduce our framework, where SHERPA is used to generate the reference jet in p+p collisions, LBT is applied to study jet evolution and medium response in hot-dense medium,  and the extend cluster hadronization model is adopted to simulate the transition of partons to hadrons.
In Sec.~\ref{sec:results}, we first present our numerical results for various $Z^0/\gamma$+hadron correlations at 5.02 TeV in both p+p and Pb+Pb collisions and compare them with experimental data, including the azimuthal angle correlations $\Delta \phi_\text{Z,ch}$,  charged hadron  transverse momentum spectra, as well as the fragmentation pattern respect to the $Z^0$ boson. Finally, we study  the  medium modification on jet substructures, such as jet shape, charged hadron number density as well as jet fragmentation functions. A short conclusion is made in Sec.~\ref{sec:conclusion}.

\section{framework }
\label{sec:framework}
\subsection{Jet production in  p+p collisions}

In our computation,   $Z^0/\gamma$-tagged jet events in  p+p  collisions are simulated with a MC event generator SHERPA2.24~\cite{Gleisberg:2008ta}, which can perform next-to-leading order (NLO) matrix element (ME) matched to the resummation of parton shower (PS) calculations with several merging schemes. AMEGIC++~\cite{Krauss:2001iv} and COMIX~\cite{Gleisberg:2008fv} are SHERPA's original matrix-element generators which provide tree-level matrix-element  and create the phase-space integration. The MC program OpenLoops~\cite{Cascioli:2011va} is interfaced with SHERPA to provide the virtual matrix elements.  The MEPS@NLO merging method~\cite{Hoeche:2009rj,Hoche:2010kg,Hoeche:2012yf} is then applied to yield an improved multi+jets production with NLO matrix elements matched to the resummation of parton showers~\cite{Gleisberg:2007md,Schumann:2007mg}. The parton distribution set``CT14 NLO"~\cite{Dulat:2015mca} is utilized to obtain the hard cross section in  p+p  collisions.

\begin{figure*}[t]
\begin{center}
\includegraphics[width=0.45\textwidth]{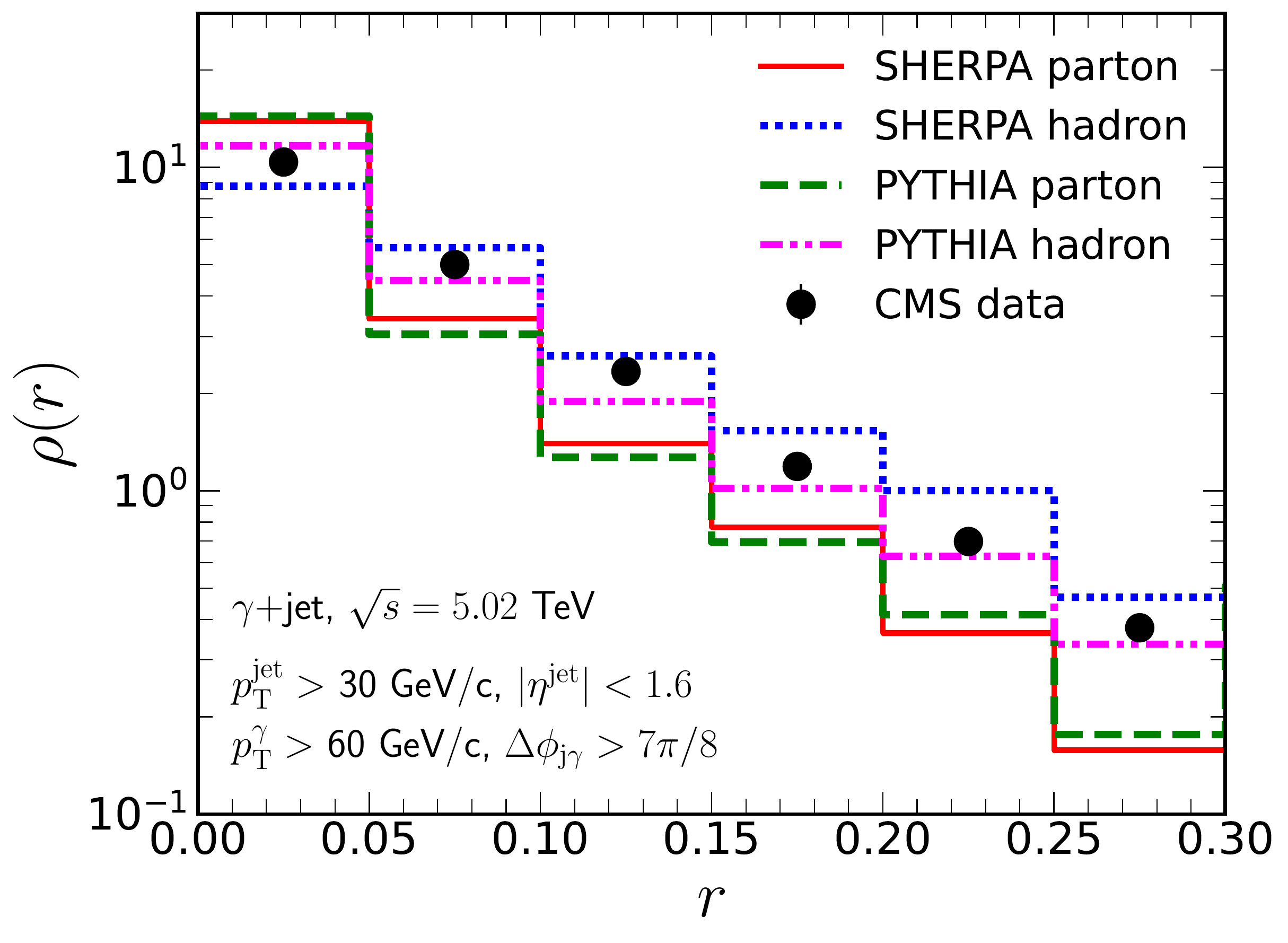}
\includegraphics[width=0.44\textwidth]{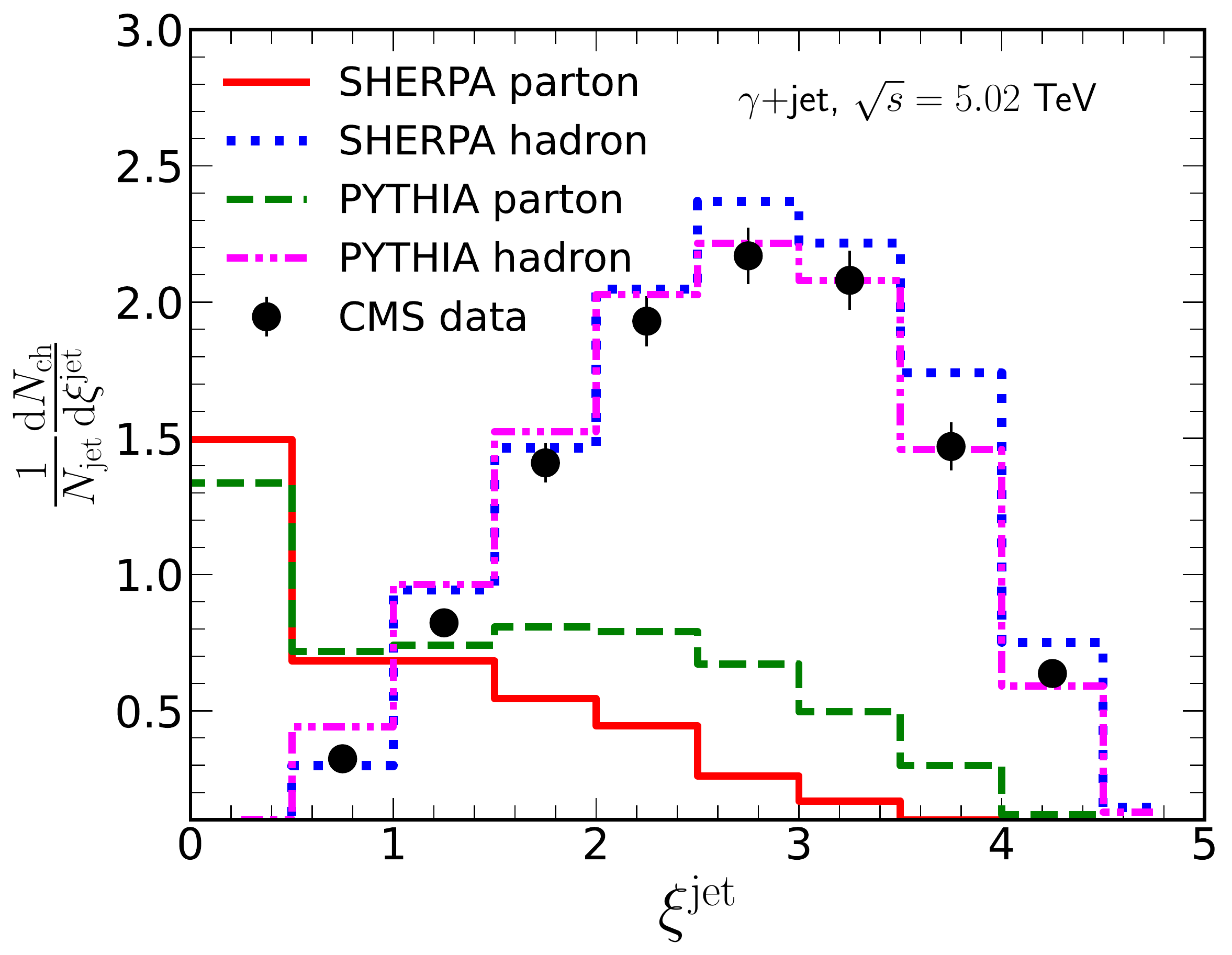}

\caption{ Jet shape (left panel) and  jet fragmentation function (right panel) are calculated at hadron level and parton level  by SHERPA  as well as by   PYTHIA and the comparison to experimental data in  p+p  collisions~\cite{CMS:2018jco,Sirunyan:2018qec}.
}
\label{jetsubtrscture}
\end{center}
\end{figure*}

SHERPA has its own fragmentation/hadronization model~\cite{Gleisberg:2008ta}, which can give nice descriptions on many experimental observables in  p+p  collisions.  Here we consider two examples of jet substructure observables. Firstly we consider jet shape, defined  as:
\begin{equation}\label{eq:jeshape}
\rho(r)=\frac{1}{\delta r} \frac{\sum_\text{jet}\sum_{r_a \leq r<r_b}(p_\text{T}^\text{ch,h}/p^\text{jet}_\text{T})}{\sum_\text{jet}\sum_{0\leq r<r_f}(p_\text{T}^\text{ch,h}/p^\text{jet}_\text{T})},
\end{equation}
  where $\delta_r=r_b-r_a $ is the width of the annulus of inner and outer radii $r_a$ and $r_b$ with respect to the jet axis, $p_\text{T}^\text{ch,h}$, $ p^\text{jet}_\text{T}$ is the transverse momentum of charged hadrons and jets respectively, $r=\sqrt{(\phi^\text{ch,h}-\phi^{\text{j}})^2+(\eta^\text{ch,h}-\eta^\text{j})^2}$ is the distance between the charged hadron and jet in the $\eta-\phi$ plane, $r_f$ is used to normalize the jet shape.  We fix $r_f=\Delta R=0.3$ to match the setup of CMS measurement~\cite{CMS:2018jco}.
In the top panel of Fig.~\ref{jetsubtrscture}, we plot jet shapes tagged with isolated photons at parton level and hadron level by SHERPA, which are also confronted with experimental data~\cite{CMS:2018jco}. It is observed that though the theoretical results with hadronization effects describe nicely CMS data in p+p~\cite{CMS:2018jco}, the results at parton level overshoot the experimental data in the the region  $ r < $0.05 and undershoot CMS data in the region  $ r > 0.05$.
This is because, jet shape is the
radial average transverse momentum distributions inside
the jet. At parton level, the jet energy is carried by few partons with large $p_\text{T}$ near the jet axis, so $\rho(r)$ peaks at $r\sim0$. However, some of jet energy will be carried  away from the jet axis by the fragmented hadrons via hadronization, leading to the enhancements of $\rho(r)$ at large $r$ compared to that at parton level.

Secondly, we investigate jet fragmentation fraction, which presents the fragmentation pattern of jet constituents with respect to the transverse momentum of the reconstructed jet and is defined as:
\begin{equation}\label{eq:jet_frag}
\xi^\text{jet}=\ln \frac{|\overrightarrow{p}^\text{jet}|^2}{\overrightarrow{p}^\text{ch,h}\cdot\overrightarrow{p}^\text{jet}}
\end{equation}
 The distributions of jet fragmentation at parton level and hadron level by SHERPA with CMS data in p+p~\cite{Sirunyan:2018qec} are shown in the bottom panel of Fig.~\ref{jetsubtrscture}.   It shows that numerical results by SHERPA at hadron level give a decent description of CMS data on the jet fragmentation distribution, whereas the simulation at parton level is a far cry from the measurement. 
 Not only the shape of jet fragmentation function  is quite different, but the normalization is radically different on parton and hadron level. This is because most of jets consist of  few partons with large $p_\text{T}$ at parton level, so the distribution of $\xi^\text{jet}$ peaks at $\xi^\text{jet}\sim 0$. At hadron level, those parton will further fragment into several hadrons, leading to the enhancement of soft particles and  the shift of $\xi^\text{jet}$ to larger value. To  confirm the hadronization effect on jet substructures, we also check the results of PYTHIA~\cite{Sjostrand:2014zea}, which give similar features as SHERPA at parton and hadron level as shown in Fig.~\ref{jetsubtrscture}. These two examples demonstrated clearly that hadronization effects are indispensable in precise calculations of some jet substructures, especially when related to particle number density and hadron constituents.

\subsection{Parton propagation in heavy-ion collisions}
When an energetic parton passing through the QGP, it may interact with the thermal medium and lose energy. In this work,  parton propagation and the corresponding medium response in hot/dense QCD medium due to jet-medium interactions are simulated by a Linear Boltzmann transport (LBT) model~\cite{Li:2010ts,He:2015pra, Cao:2016gvr}, based on a Boltzmann equation:
\begin{equation}\begin{split}
 p_1\cdot\partial f_a(p_1)&=-\int\frac{\text{d}^3p_2}{(2\pi)^32E_2}\int\frac{\text{d}^3p_3}{(2\pi)^32E_3}\int\frac{\text{d}^3p_4}{(2\pi)^32E_4}\\
&\sum _{b(c,d)}[f_a(p_1)f_b(p_2)-f_\text{c}(p_3)f_d(p_4)]|M_{ab\rightarrow cd}|^2\\
&\times S_2(s,t,u)(2\pi)^4\delta^4(p_1+p_2-p_3-p_4)
 \end{split}\end{equation}
 $f_i$ are phase-space distributions of partons.
 $S_2(s,t,u)$ is Lorentz-invariant regulation condition to regulate all soft and collinear  divergence.
 Elastic scattering is introduced by the corresponding $|M_{ab\rightarrow cd}|$ which includes the complete set of leading order  2-2 elastic scattering processes~\cite{He:2015pra}.

The induced gluon radiation from inelastic scattering is numerically incorporated into LBT based on High-Twist formalism~\cite{Guo:2000nz,Zhang:2003yn,Zhang:2003wk}:

\begin{equation}
\frac{\text{d}N_\text{g}}{\text{d}x\text{d}k_\perp^2 \text{d}t}=\frac{2\alpha_sC_AP(x)\hat{q}}{\pi k_\perp^4}\left(\frac{k_\perp^2}{k_\perp^2+x^2M^2}\right)^4\sin^2\left(\frac{t-t_i}{2\tau_\text{f}}\right)
 \end{equation}
 in which, $x$ and $k_\perp$ give the energy fraction and transverse momentum of the radiated gluon respectively,
 $ P(x)$ denotes splitting functions.  $\hat{q}$ stands for the jet transport coefficient and is extracted from the elastic scattering.
 $\tau_\text{f}=2Ex(1-x)/(k_\perp^2+x^2M^2) $ characterizes the formation time of the radiated gluon. The medium information is provided by 3+1D CLVisic hydrodynamics~\cite{Pang:2012he,Pang:2014ipa} and the initial condition is provided by AMPT~\cite{Lin:2004en}.
 LBT has been successful in describing the suppression of experimental data on hadrons~\cite{He:2015pra, Cao:2016gvr}, inclusive jets~\cite{He:2018xjv}, $\gamma+$hadron/jets~\cite{Chen:2017zte,Luo:2018pto} correlations and $Z^0$+jet production~\cite{Zhang:2018urd}.

The original version of LBT~\cite{Li:2010ts,He:2015pra, Cao:2016gvr}  have an excellent description of the momentum exchange between jet shower partons and hot dense medium. In fact,  jet shower partons not only exchange momentum with the medium partons, but also exchange color-charge with the medium constituents  during its propagation in the medium.
To have the whole evolution history information  of a event in Pb+Pb as that in  p+p  collisions, we future tracked the color flow of the event during its propagation in the medium based on LBT model. The color index of shower partons  can be  read from  SHERPA ( or other event generators). Meanwhile, we sample  the color  index of the thermal partons randomly.  Then we track the color flow as illustrated in~\cite{Boos:2001cv} and in the bottom panel of Fig.~\ref{illutration} in elastic scattering by identifying all types of gluon lines in each elastic process according to the flavor of the incoming and outgoing partons. Furthermore, we also keep track of the color flow in gluon radiation process.

\subsection{Hadronization in p+p and A+A}
 Once the partons escape the medium or the temperature fall below the transition temperature, they will quickly hadronize into color neutral bound states. In this section, we will present an extended cluster model of hadronization, which generalizes the Cluster model for p+p reactions in SHERPA to consider hadronization processes in heavy-ion collisions.

 The QCD Cluster Model was conceptually  proposed and developed by  References~\cite{Gottschalk:1982yt,Gottschalk:1983fm,Gottschalk:1986bv,Webber:1983if}. For p+p collisions, the non-perturbative transition of these partons into primary hadronic matter is accomplished following  the procedure of reference~\cite{Webber:1983if} and is summarized as: 

    \begin{figure}[t]
    \begin{center}
    \includegraphics[width=0.5\textwidth]{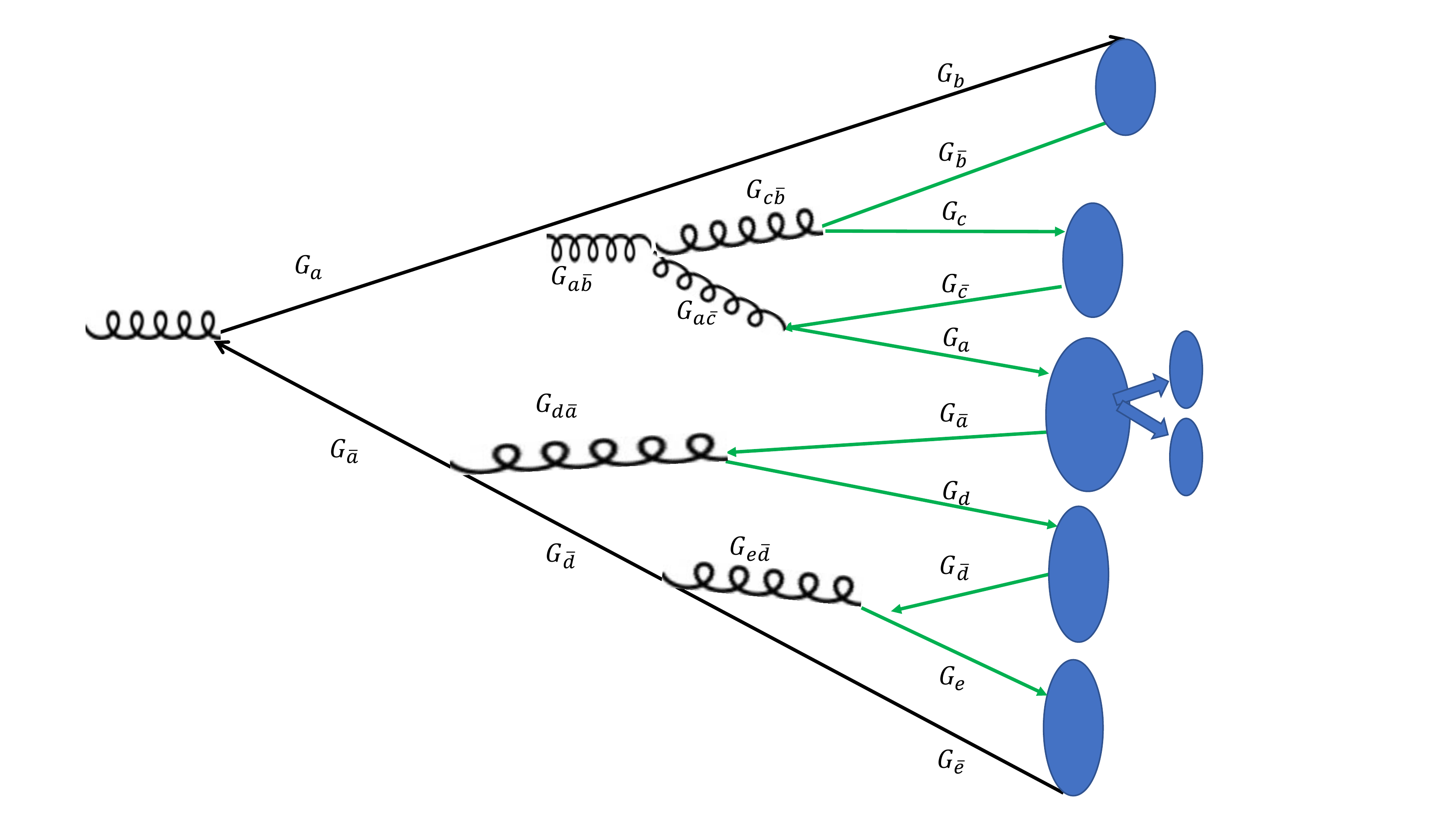}
    \includegraphics[width=0.5\textwidth]{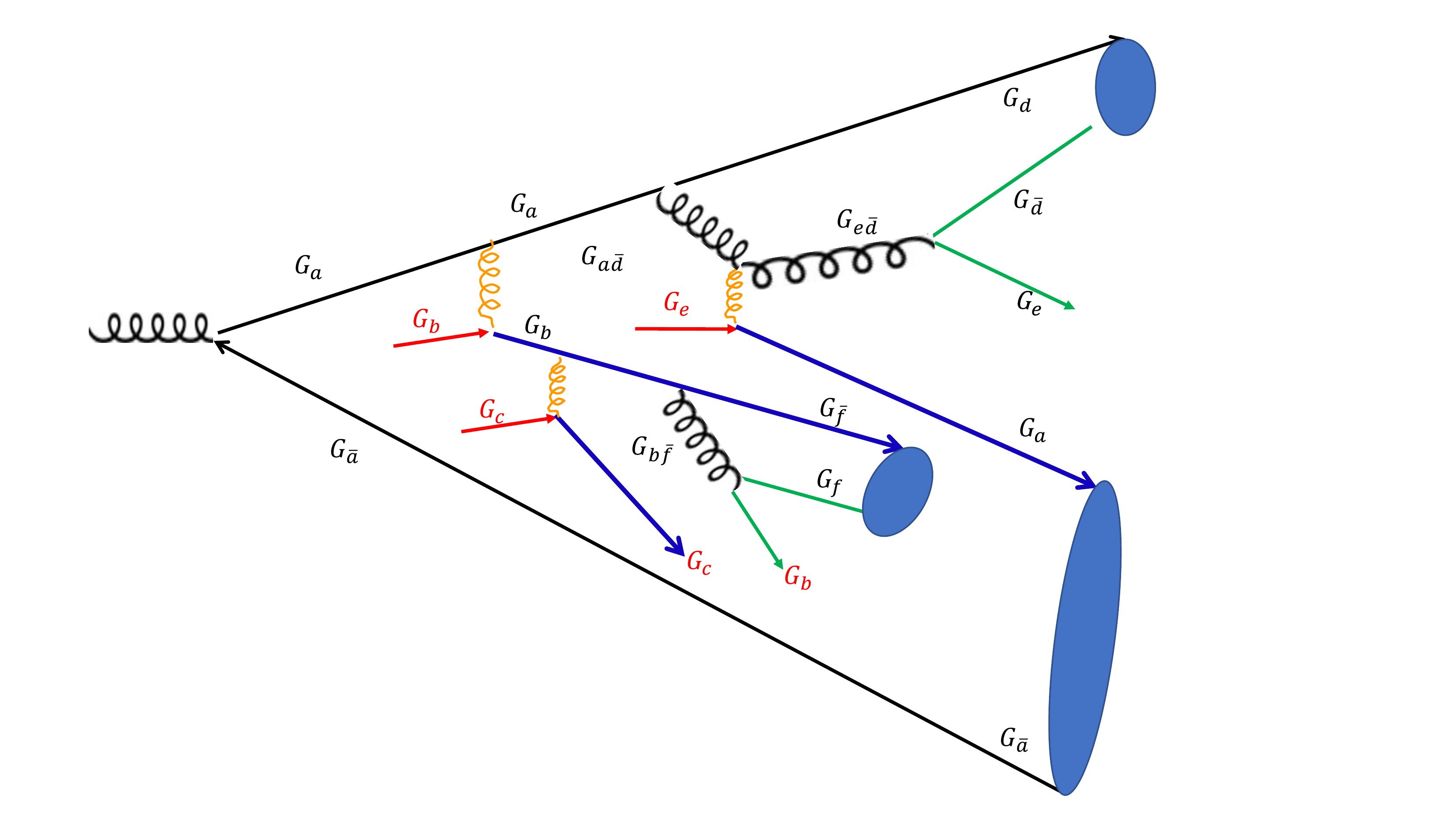}
    \caption{ Schematic illustration of the procedure used to isolate color singlet clusters in p+p (top) and A+A (bottom) collisions. $G_i$ is the color index  carried by the parton.
    }
    \label{illutration}
    \end{center}
    \end{figure}

    \begin{itemize}
      \item The shower partons from SHERPA are massless
because they are considered as final status particles. However, only with mass can those partons hadronize into color neutral hadrons. 
We chose quark masses according to references~\cite{Gottschalk:1982yt,Gottschalk:1983fm,Gottschalk:1986bv,Webber:1983if}: $ M_\text{(u,d)}=0.3  $ GeV/$c^2$, $ M_\text{s}=0.45  $ GeV/$c^2$, $ M_\text{c}=1.5 $ GeV/$c^2$, $ M_\text{b}=5 $ GeV/$c^2$, while keeping the parton momentum unchanged.

      \item
           Clusters are formed by quark pairs, so gluons split into  $q\bar{q}$ pair following the probability (splitting) function $f(z)=z^2+(1-z)^2$, where $z$ denotes the energy fraction carried by fragmented quarks. The color flow of gluon is shared by $q\bar{q}$ correspondingly. 

      \item  All final partons are color connected, the color connected quark-antiquarks are combined to form a color neutral cluster which is uncorrelated to other clusters  and can be seen as an independent string~\cite{Webber:1983if}. Schematic illustration of the procedure used to isolate color singlet is shown in Fig.~\ref{illutration}.

      \item For a cluster C with invariant mass $M_\text{C} > M_\text{f}$, where $M_\text{f}$ is a parameter and  need to be fixed via comparison with experimental data, formed from a (di)quark of momentum $p_1$ and an anti(di)quark of momentum $p_2$, the fission C $\rightarrow$ X + Y is assumed to yield~\cite{Webber:1983if}
          \begin{equation}\begin{split}
          &p_x=(1-x)p_1+xp_2 \\
          &p_y=(1-x)p_2+xp_1
         \end{split}\end{equation}
         $x$ is sampled according to the distribution $f(x)=x^2+(1-x)^2$.
         The flavour of the produced pair is taken to be $q\bar{q}$ and $D\bar{D}$, where $q$ and $D$ represent quark and di-quark.  
         If necessary the process is repeated for the fission fragments X and Y, until all cluster invariant mass are below the fission threshold ~\cite{Webber:1983if}.

      \item The colour neutral clusters can decay independently into hadrons~\cite{Webber:1983if}.  The cluster which consists of $q_i\bar{q_j}$ or $q_i\bar{D_j}$ will be combined with  $q_{i'}\bar{q_{j'}}$, $D_{i'}\bar{D_{j'}}$ coming from vacuum to form hadrons:  $q_i\bar{q_j'}\ \ q_{i'}\bar{q_j}$, $q_i\bar{D_{j'}}\ \ D_{i'}\bar{q_j}$, $q_i\bar{q_{j'}}\ \ q_{i'}\bar{D_j}$,   $q_i\bar{D_{j'}}\ \ D_{i'}\bar{D_j}$.  The PDG of the hadron is determined by the flavor and spin of the daughter quarks (sampled by hand), as well as the mass of the cluster and  the candidate hadron.  At present, only spin 1/2 and 3/2 baryon and spin 0 and 1 meson are considered in our simulation.  The energy and momentum of the final hadrons are the solution of the energy-momentum conservation equation.

      \item And then, all the unstable hadrons will decay into final stable particles according to its properties: branching ratio, number of decay products, and PDG of decay produces etc... If necessary the process is repeated  until all final particles become stable.

    \end{itemize}

\begin{figure}[t]
\begin{center}
\includegraphics[width=0.45\textwidth]{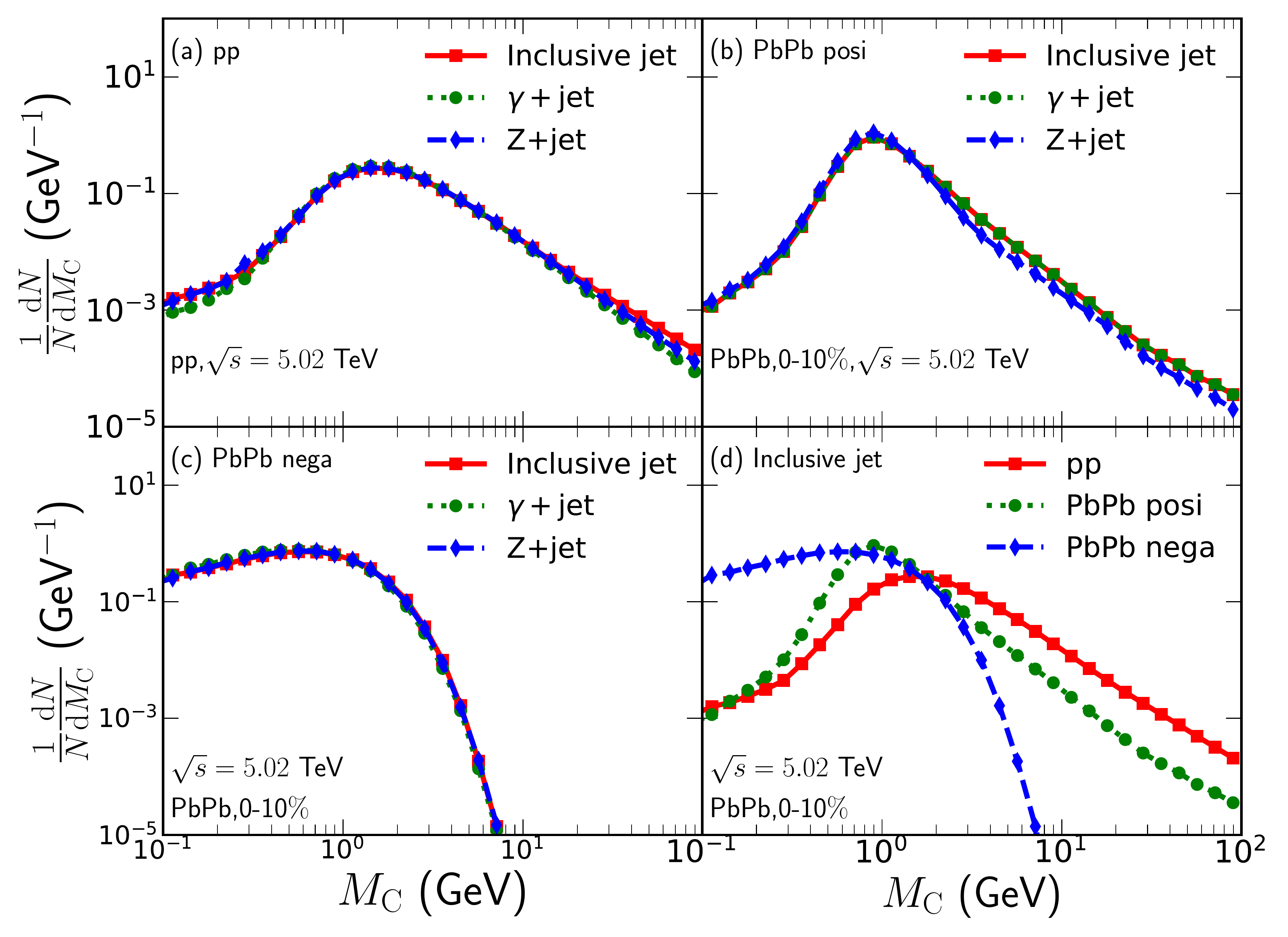}
\vspace{-10pt}
\caption{ Distribution of cluster invariant mass: (a) in p+p, (b) of positive in 0-10$\%$ Pb+Pb, (c) of negative in 0-10$\%$ Pb+Pb collisions  of inclusive jet, $\gamma$+jet and $Z^0$+jet process at $\sqrt{s}=5.02$ TeV.  (d) is the comparison of $M_\text{C}$ distribution  between pp, positive and negative of inclusive jet. 
}
\label{MC_mass}
\end{center}
\end{figure}

        In Pb+Pb collisions,  jet shower partons will exchange color-charge with the medium constituents.  As a consequence, some jet shower  partons and jet induced thermal partons may carry the color index which is carried originally by the initial  thermal parton, therefore, those partons are color connected with medium partons (opposite color index of initial thermal partons)  rather than other shower partons as in p+p collisions due to elastic scatterings as illustrated in the bottom picture of Fig.~\ref{illutration}.  Meanwhile, initial  thermal partons are also color connected to other medium constituents with opposite color index.  In the following, jet shower  partons and  recoil partons are called {\it positive partons}, while thermal partons referred to  {\it negative partons}  as defined in LBT~\cite{He:2015pra}. Schematic illustration of the procedure used to isolate color singlet in Pb+Pb collisions is shown in the bottom panel of Fig.~\ref{illutration}.

    We simulate the hadronization in heavy-ion collisions with the extended cluster model as:
     \begin{itemize}
      \item
    We combine the positive (denoted as ``posi" in the figures)  and negative (denoted as ``nega" in the figures)  quarks into clusters {\it separately}, similar as JETSCAPE~\cite{JETSCAPE:2022jer}.
   Considering the complexity and technicality of the positive-negative case, we neglect the contributions from mixture of positive partons with negative partons at present.

    \item Those positive partons, which have positive color partners will be treated in the same approach as in   p+p  collisions with the original Cluster model~\cite{Webber:1983if}.  In the mean time,  those positive partons that do not have positive color partners (color connected to negative partons) may be combined into clusters with each other  according to the distance $r=\sqrt{(\Delta \phi)^2+(\Delta \eta)^2}$ between them.    

    \item A negative parton is requested to form a cluster with its closest negative parton. As a consequence, those constructed clusters, whose daughter partons are not color connected to its anti-color partners (the later two cases),  may  be not color neutral, different from the color neutral clusters in p+p collisions.   Later,  these clusters will radiate some soft gluons to become color neutral.

    \item The four-momentum of the cluster is the sum of the daughter partons.

 \item
 The negative parton distributions will be subtracted form positive parton distributions within the kinematic region they appear (denoted as ``posi+nega" in the figures), as treated in LBT~\cite{Li:2010ts,He:2015pra, Cao:2016gvr}.

  \end{itemize}

    In Fig.3, we present normalized distributions of the cluster invariant mass $M_\text{C}=\sqrt{E_\text{C}^2-p_\text{C}^2}$: (a) in p+p, (b) of positive in 0-10$\%$ Pb+Pb, (c) negative in 0-10$\%$ Pb+Pb collisions   for three different processes, i.e., inclusive jet and $\gamma $+jet as well as Z+jet at $\sqrt{s}=5.02$ TeV.
It is illustrated that
 the $M_\text{C}$ distributions for three processes in p+p (or Pb+Pb) overlap with each other, and show negligible dependence on the processes both in p+p and Pb+Pb. Therefore, this method is universal. 
The comparison of $M_\text{C}$ distribution  between p+p, positive and negative of inclusive jet are also shown in Fig.3 (d).

 The $M_\text{C}$ distributions of positive partons in A+A are shifted to smaller value as compared to p+p, which results from two underlying reasons. First, the  distance of the color connected positive partons will  become closer in Pb+Pb collisions due to medium-induced radiation. Some jet shower partons tend to  be connected to its radiated gluon as illustrated  in  the bottom panel of Fig.~\ref{illutration}. Second, the clusters formed according to the distance  will also have smaller $M_\text{C}$. The $M_\text{C}$ distributions of negative partons in A+A spread mainly in small $M_\text{C}$ region because of their much lower energies.

It is noted that the cluster with invariant mass $M_\text{C} > M_\text{f}$ may split~\cite{Webber:1983if} and it is found that results with $M_\text{f}=3.0 $~GeV can excellently  describe the experimental data of  the correlations  between $Z^0$ boson and its tagged charged hadrons~\cite{CMS:2021otx},  and $\gamma$-tagged jet substructures~\cite{CMS:2018jco,Sirunyan:2018qec} in p+p collisions. In principle, the parameter $M_\text{f}$
for positive partons in Pb+Pb collisions could be different from the one in p+p collisions due to their different
$M_\text{C}$ distributions. Considering the facts that $M_\text{f}$ is strongly correlated with the strength of jet-medium interactions in A+A collision, we should tune $M_\text{f}$ for positive and negative parton via global analysis. In this study, we fix $\alpha_s$ to be $\alpha_s$ = 0.20 in LBT as used for the study of $Z^0$-jet correlations in Pb+Pb collisions~\cite{Zhang:2018urd}; and we find that $M_\text{f}=3$ GeV can also well describe charged hadron spectrum~\cite{CMS:2021otx} at large $p_\text{T}$ as shown in Fig.6.
The only varying parameter in our extended cluster model of hadronization is $M_\text{f}$ for negative parton, which is chosen to
be 0.25 GeV to  give optimal
descriptions of experimental data of the azimuthal angle correlation between charged hadron and the recoiling $Z^0$ boson~\cite{CMS:2021otx}, which is sensitive to  hadrons from negative partons, in heavy-ion collisions as shown in Fig.4.

\section{Numerical Results}
\label{sec:results}
\subsection{Parton-medium interaction effect on $Z^0/\gamma$-hadron correlations }

 With the framework described in Sec.~\ref{sec:framework}, we present a detailed study of $Z^0$-hadron correlation  as well as hadron spectra tagged with $Z^0$ boson in both  p+p  and Pb+Pb collisions.
In order to compare with experimental measurements, we select the $Z^0$ boson in association with jets according to the kinematic cuts adopted by CMS  experiment~\cite{CMS:2021otx}.  To compare with CMS results, the electrons are constrained in the phase space $p_\text{T} > 20$ GeV/$c$ and $|\eta| <2.1$ and are rejected in the transition region $1.44 < |\eta| < 1.57$; muons are required to have $p_\text{T} >20 $ GeV/$c$ and $|\eta|<2.4$; the $Z^0$ candidates are defined as electron or muon pairs, with a reconstructed invariant mass ($M^{ll}$) in the phase space $60-110$ GeV/$c^2$ and $p_\text{T}^Z>$30 GeV/$c$. 
The charged hadrons are  required in the rapidity region $|\eta^\text{ch,h}|<2.4$ and in the transverse momentum region $p_\text{T}^\text{ch,h}>1$ GeV/$c$.

\begin{figure}[t]
\centering
\includegraphics[width=0.45\textwidth]{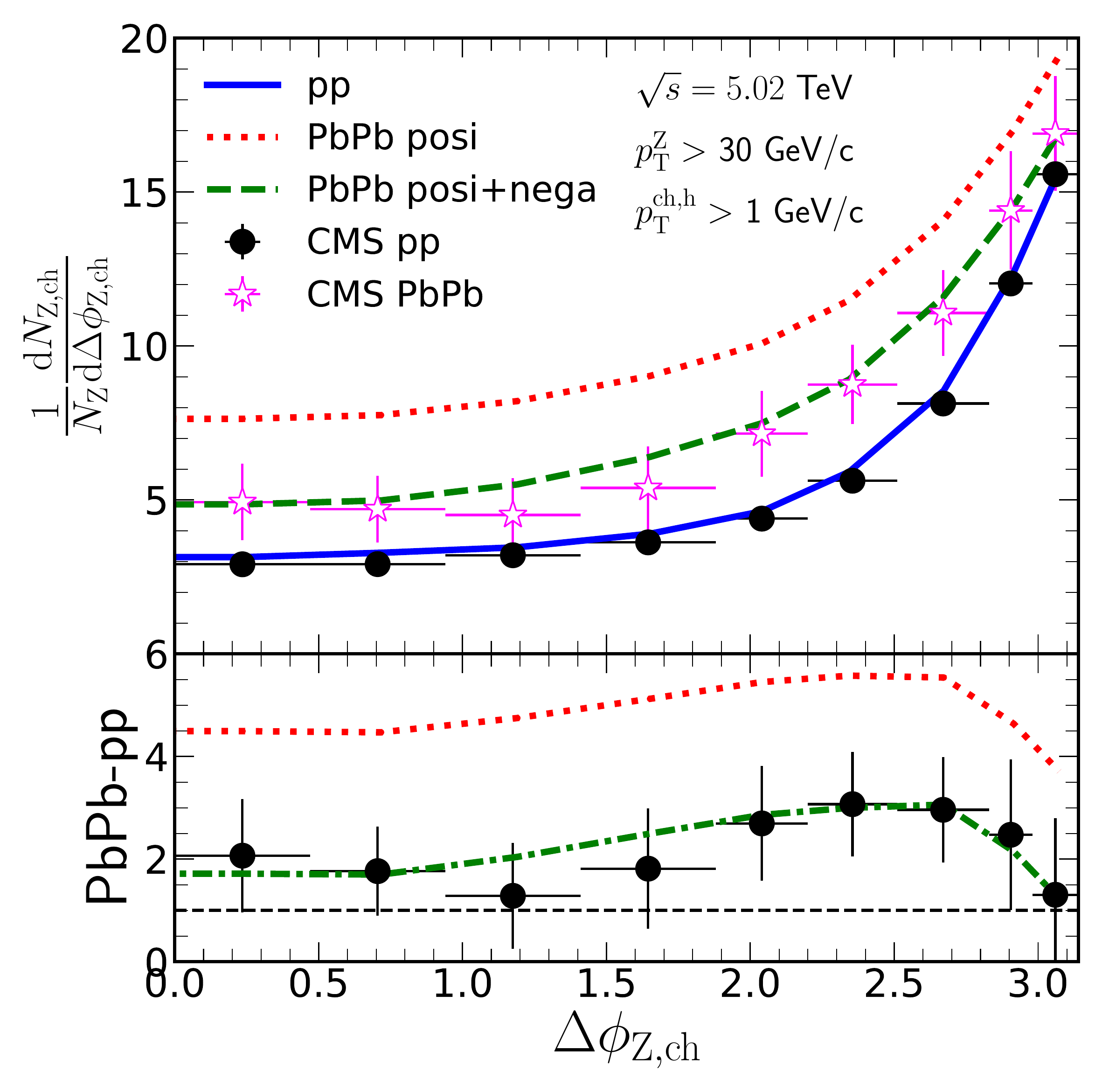} \\
\vspace{-0pt}
  \caption{(Color online) Top: the azimuthal  angle correlation $\Delta \phi_\text{Z,ch}$ between  charged hadrons and  the recoiling $Z^0$ boson in   p+p  and  Pb+Pb collisions with experimental data~\cite{CMS:2021otx}. Bottom: the difference of the $\Delta \phi_\text{Z,ch}$  distribution between  Pb+Pb and   p+p  collisions. } \label{phi_zh}
\end{figure}

Firstly, the distributions of the azimuthal angle difference $\Delta \phi_\text{Z,ch}=|\phi_Z-\phi_\text{ch,h}|$ between charged hadrons and the recoiling $Z^0$ bosons, normalized by the number of $Z^0$ bosons both  in  p+p  and  Pb+Pb collisions  as well as the
comparison  with  CMS measurements data~\cite{CMS:2021otx} are shown in the top panel of  Fig.~\ref{phi_zh}. Excellent agreements are obtained for $\Delta \phi_\text{Z,ch}$ distributions in both   p+p  and Pb+Pb collisions.
Our calculations with only hadrons from positive partons  overshoot the measured distribution of  $\Delta \phi_\text{Z,ch}$ in Pb+Pb collisions. However, when hadronization effects from negative partons are taken into account, our result presents a very nice description of  the experimental data in Pb+Pb. 
 Compared to  p+p  collisions, a significant enhancement of  $\Delta \phi_\text{Z,ch}$ distribution in Pb+Pb collisions is  observed, which results from the fact that the $\Delta \phi_\text{Z,ch}$ distribution is dominated by soft particles, and the increasing of soft hadron yields due to jet-medium interactions lead to the overall enhancement of $\Delta \phi_\text{Z,ch}$ in Pb+Pb collisions as discussed in more details later.

 To quantify the relative shift of $\Delta \phi_\text{Z,ch}$ in Pb+Pb collisions due to jet-medium interactions, we show the difference of $\Delta \phi_\text{Z,ch}$  between  p+p  and Pb+Pb collisions in bottom panel of Fig.\ref{phi_zh}. The distribution of $\Delta \phi_\text{Z,ch}$ is enhanced  with the largest magnitude at $\Delta \phi_\text{Z,ch} \sim 2.5$ region, while with the smallest magnitude at $\Delta \phi_\text{Z,ch}\sim \pi $, resulted from the reduction of large $p_\text{T}$ hadrons, produced in the opposite direction of $Z^0$ boson,  as explained in the following.
 Besides,  the beam remnants~\cite{Sjostrand:1987su,Beraudo:2022dpz}, the remnant of the projectile  moving in the same direction as beam with forward rapidity, which have significant contributions to the soft hadron yields, especially for soft hadrons in the same-side ($\Delta \phi_\text{Z,ch}<\pi/2$) region, should also be included. Because the beam remnants would be color connected to jet partons, the cluster mass would be very large and they will experience huge amount of string splitting processes.  As a consequence, they would contribute significantly to soft hadrons yields in the whole region.

\begin{figure}[t]
\begin{center}
\vspace{5pt}
\includegraphics[width=0.45\textwidth]{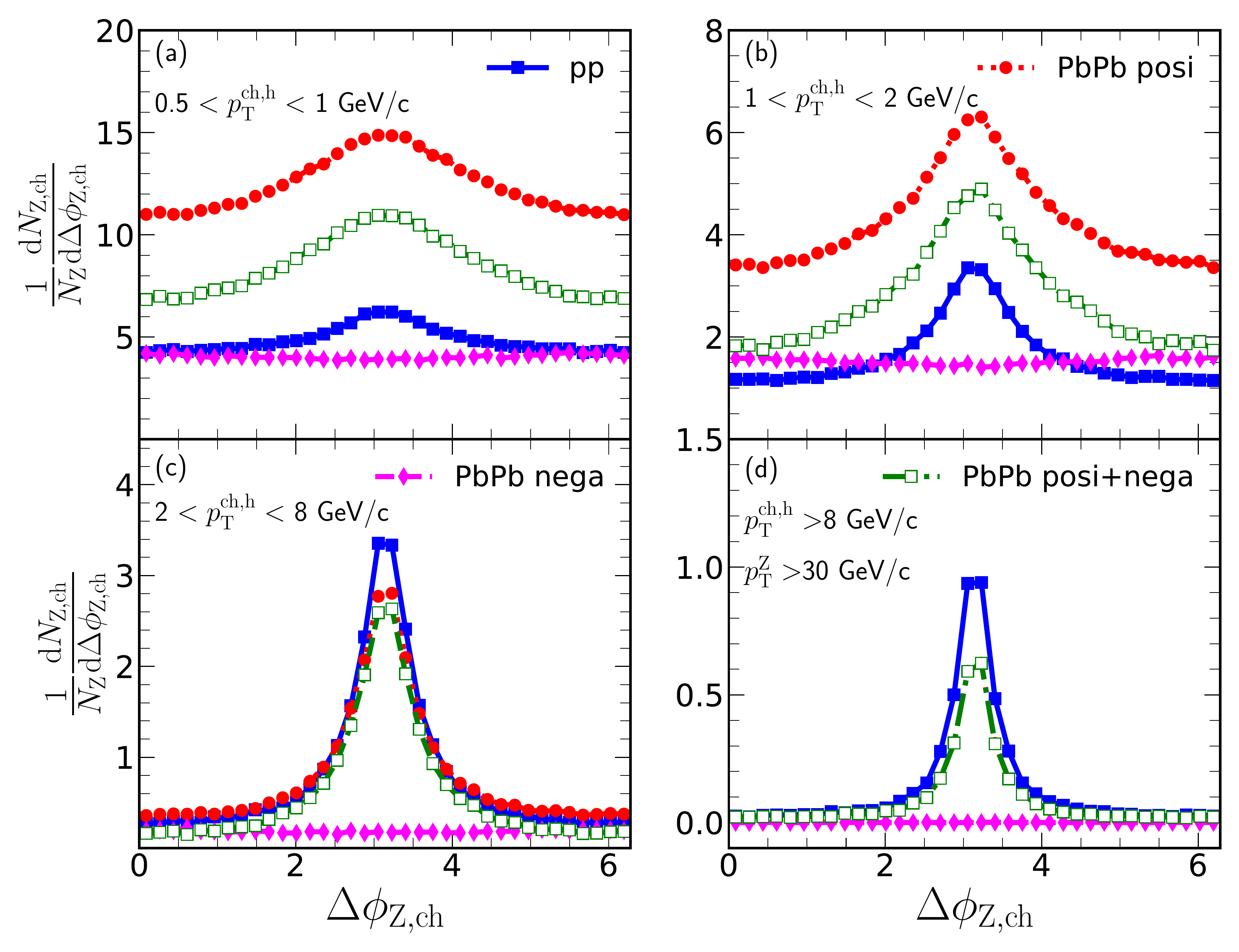}
\end{center}
\vspace{-10pt}
\caption[*]{(Color online) The distributions of the azimuthal angle difference $\Delta \phi_\text{Z,ch}$ between charged particles from positive (filled circle), negative (filled diamond) as well as positive minus negative (circle, denoted as ``posi+nega" ) and the recoiling $Z^0$ bosons, normalized by the number of $Z^0$ bosons in different $p_\text{T}^\text{ch,h}$ regions in  p+p  and Pb+Pb collisions. }

\label{phi_zh_pt}
\end{figure}
 
To have a detailed understanding of the modification of the azimuthal angle correlation, we calculated the $\Delta \phi_\text{Z,ch}$ distributions in four $p_\text{T}^\text{ch,h}$ intervals: 0.5-1 GeV/$c$, 1-2 GeV/$c$, 2-8 GeV/$c$, $>$8 GeV/$c$ both in  p+p  and Pb+Pb collisions and show the results in Fig.~\ref{phi_zh_pt}.  The contributions from negative partons and positive partons to $\Delta \phi_\text{Z,ch}$ are also calculated.
 We find that contributions from negative partons hadronization are uniformly distributed in the whole phase space and are independent on the $\Delta \phi_\text{Z,ch}$, because negative
partons are randomly sampled in the medium~\cite{Li:2010ts,He:2015pra, Cao:2016gvr}.  On the the hand, the hard partons are produced at the opposite direction of $Z^0$ boson in reference p+p collisions, as a result of which,  those from positive partons hadronization are Gaussian distributed with mean value $\Delta \phi_\text{Z,ch}=\pi$. It is shown that contributions from negative partons  are essential for soft hadron yields, but become negligible when high $p_\text{T}^\text{ch,h}$ hadrons are concerned.  In $p_\text{T}^\text{ch,h} > 2 $ GeV/$c$ region, the distributions of  $\Delta \phi_\text{Z,ch}$ are overall suppressed in Pb+Pb collisions (with or without contribution of negative partons hadronization) relative to p+p, while $p_\text{T}^\text{ch,h}<2 $ GeV/$c$ they are enhanced in Pb+Pb collisions due to the effect of jet quenching and hadronization. 
Therefore, the $\Delta \phi_\text{Z,ch}$ distribution is dominated by soft particles, and the increasing of soft hadron yields lead to the overall enhancement of $\Delta \phi_\text{Z,ch}$ in Pb+Pb collisions as shown in Fig.4. Besides, the reduction of large $p_T$ hadrons, produced in the opposite direction of $Z^0$ boson, lead to the smallest enhancements at $\Delta \phi_\text{Z,ch} \sim \pi$.

\begin{figure}[t]
\centering
\includegraphics[width=0.45\textwidth]{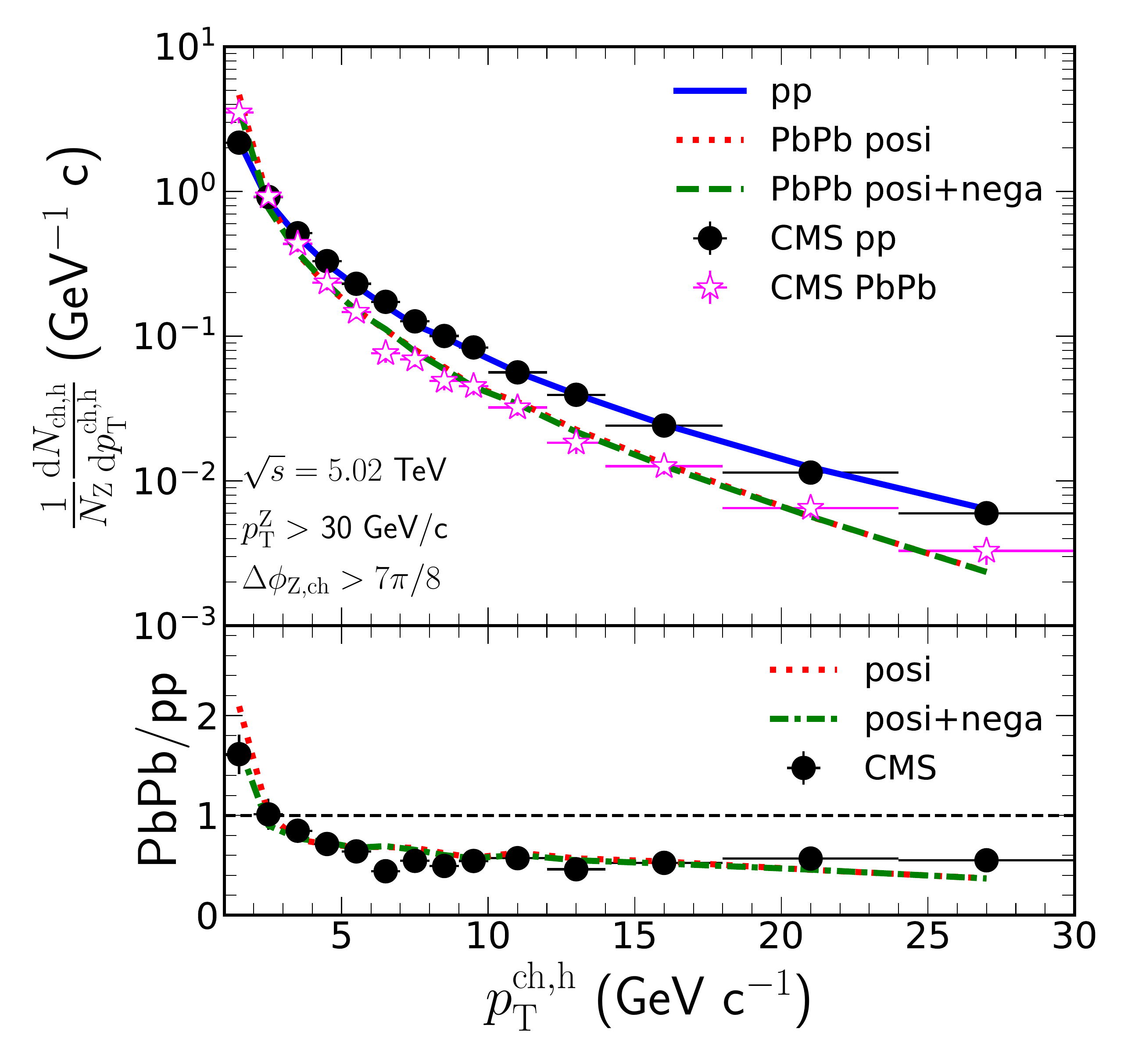} 
\vspace{-5pt}
  \caption{(Color online) Top: Comparison between our model calculations of $Z^0$ boson tagged charged hadron $p_\text{T}^\text{ch,h}$ distributions  in  p+p  and Pb+Pb collisions with CMS experimental data~\cite{CMS:2021otx}. Bottom: the ratio of the $p_\text{T}^\text{ch,h}$  distributions in Pb+Pb to that in p+p collisions. } \label{Zh_pt}
\end{figure}

In the top panel of Fig.~\ref{Zh_pt}, we present the charged hadron transverse momentum $p_\text{T}^\text{ch,h}$ spectra in p+p and Pb+Pb collision and compare them to CMS experimental data~\cite{CMS:2021otx}. In the simulation, the charged hadrons are required to satisfy $\Delta \phi_\text{Z,ch}> 7\pi/8$. Our calculations are in good agreement with the experimental data~\cite{CMS:2021otx} in both p+p and Pb+Pb collisions, while numerical results only from positive parton in Pb+Pb collisions may overestimate the soft hadron ($p_\text{T}^\text{ch,h}<2$ GeV/$c$). 
The ratio of charged hadron $p_\text{T}$ distribution in Pb+Pb collisions to that in p+p collisions is compared to experimental data~\cite{CMS:2021otx} in the bottom panel of Fig.~\ref{Zh_pt}.
We see that the particle production in Pb+Pb is suppressed at  high $p_\text{T}^\text{ch,h}$, while enhanced  at low $p_\text{T}^\text{ch,h}$ (1  $<\ p_\text{T}^\text{ch,h}\ <$ 2 GeV/$c$), as compared to the p+p reference and in consistent with the illustration  of  Fig.~\ref{phi_zh_pt}. The $p_\text{T}$ of hadrons from negative partons is small, so they only have considerable contributions to $p_\text{T}^\text{ch,h}$ spectra in low $p_\text{T}$. The $p_\text{T}$ spectrum at large $p_\text{T}$ is dominated by hadrons from positive partons, and can be used to tune the parameter $M_\text{f}$
for positive partons in Pb+Pb collisions.  

\begin{figure}[t]
\vspace{-5pt}
\centering
\includegraphics[width=0.45\textwidth]{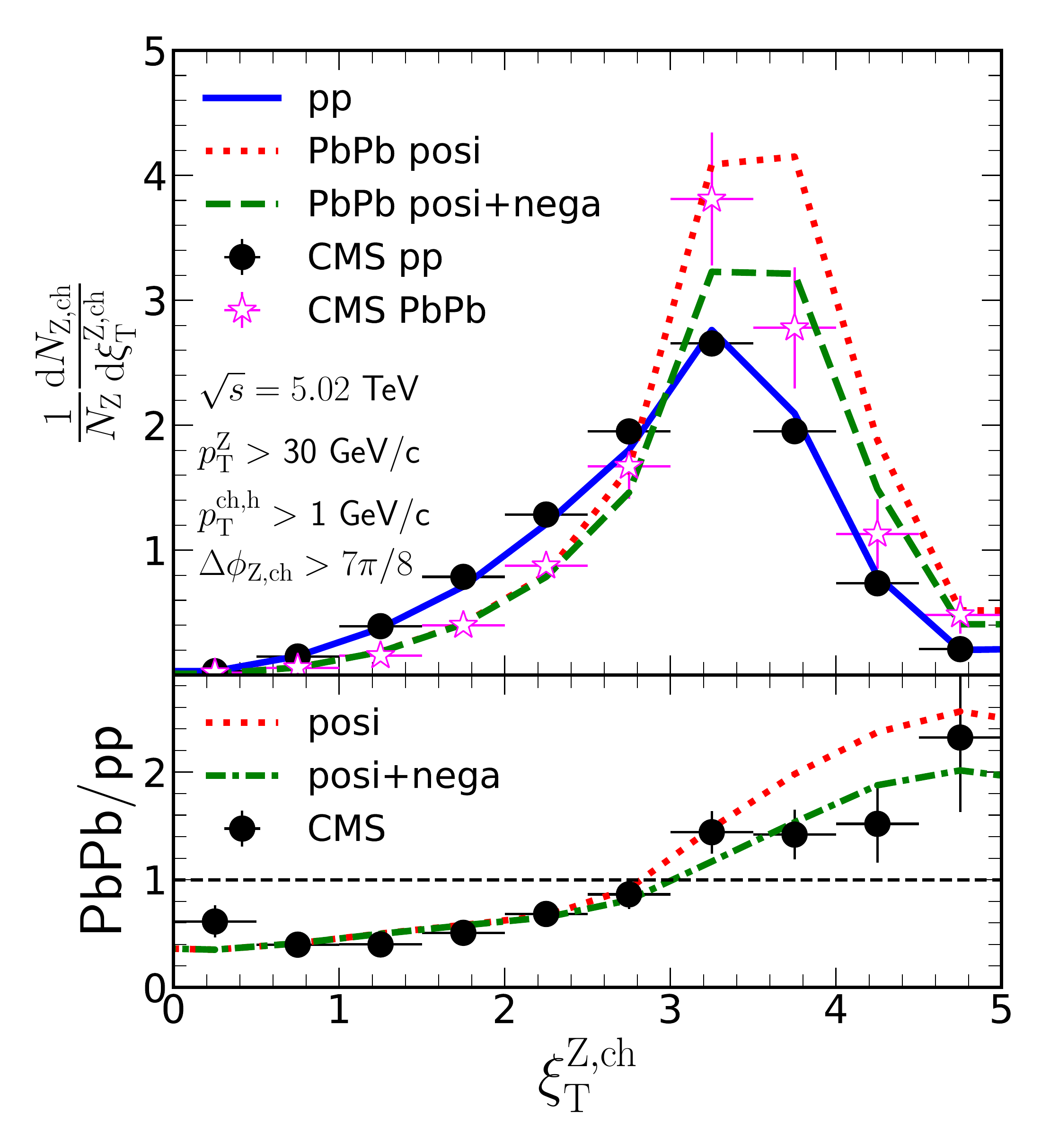}  \\
\vspace{-5pt}
  \caption{(Color online) Top :  Theoretical results of $\xi_\text{T}^\text{Z,ch}$ distributions in p+p and Pb+Pb collisions, confronted against  CMS experimental data~\cite{CMS:2021otx}. Bottom: the ratio of $\xi_\text{T}^\text{Z,ch}$ distributions in Pb+Pb to that in p+p collisions. } \label{frag_zh}
\end{figure}


Recently the fragmentation pattern  of the particles recoiling from a $Z^0$ boson,
\begin{equation}
\xi^\text{Z,ch}_\text{T}=\ln [-|\overrightarrow{p}^{Z}_\text{T}|^2/(\overrightarrow{p}^\text{ch,h}_\text{T}\cdot\overrightarrow{p}^{Z}_\text{T})]
\label{eq:xi_C}
\end{equation}
has been measured in Pb+Pb collisions by CMS Collaboration~\cite{CMS:2021otx}. The results of $\xi^\text{Z,ch}_\text{T}$ distributions in our model and their comparison with experimental data~\cite{CMS:2021otx}  are illustrated in the top panel of Fig.~\ref{frag_zh}. The ratio of $\xi^\text{Z,ch}_\text{T}$ in Pb+Pb to that in  p+p  collisions  are shown in the bottom panel of Fig.~\ref{frag_zh}.  To match the corresponding experiment kinematics~\cite{CMS:2021otx}, charged hadrons are required to satisfy $\Delta \phi_\text{Z,ch}> 7\pi/8$ and $p_\text{T}^\text{ch,h}>1 $ GeV/$c$.   We see our simulations for both p+p  and Pb+Pb collisions are  consistent with the experimental data~\cite{CMS:2021otx}.  The definition of $\xi^\text{Z,ch}_\text{T}$ in Eq.~\ref{eq:xi_C} tells that small $\xi^\text{Z,ch}_\text{T}$  region corresponds to high energy hadrons and large $\xi^\text{Z,ch}_\text{T}$ region comes from soft hadrons.  A significant suppression in small $\xi^\text{Z,ch}_\text{T}$ region and a large enhancement in large $\xi^\text{Z,ch}_\text{T}$ region are obtained in Pb+Pb collisions relative to  p+p  collisions, a trend in nice agreement with CMS data~\cite{CMS:2021otx}. We notice that the contribution from negative parton to $\xi^\text{Z,ch}_\text{T}$ is significant in large $\xi^\text{Z,ch}_\text{T}$ region (with small $p^\text{ch,h}_T$).
Including contributions from medium response, our calculations can better describe the experiment for soft hadron productions. 

We emphasize that  traditional jet fragmentation $\xi^\text{jet,ch}_\text{T}$, defined as $\ln [|\overrightarrow{p}^\text{jet}|^2/(\overrightarrow{p}^\text{ch,h}\cdot\overrightarrow{p}^\text{jet})]$ (see Eq. 2), is explicitly  dependent on jet $p_\text{T}$. On the other hand,  $\xi^\text{Z,ch}_\text{T}$, defined as $ \ln [-|\overrightarrow{p}^\text{Z}_\text{T}|^2/(\overrightarrow{p}^\text{ch,h}_\text{T}\cdot\overrightarrow{p}^\text{Z}_\text{T})]   $ (see Eq. 6), depends on $p_\text{T}^{Z}$. 
In heavy-ion collisions,
$p^\text{jet}$ and  $p^\text{ch,h}$  will be altered due to jet-medium interactions, while $p_\text{T}^{Z}$ almost keep unmodified when propagating through the QGP.  So it is understandable that the shift of the $\xi^\text{Z,ch}_\text{T}$ in Pb+Pb will be more pronounced with a better sensitivity to parton energy loss and medium response.

\subsection{Medium-Induced Modifications of Jet substructure }

We now turn to medium effects on full jet observables, and investigate the redistribution of jet substructure associated with
a $Z^0/\gamma$ in Pb+Pb relative to the reference  p+p  collisions by including hadronization effects with the extended cluster model.
\begin{figure}[t]
\centering
\includegraphics[width=0.45\textwidth]{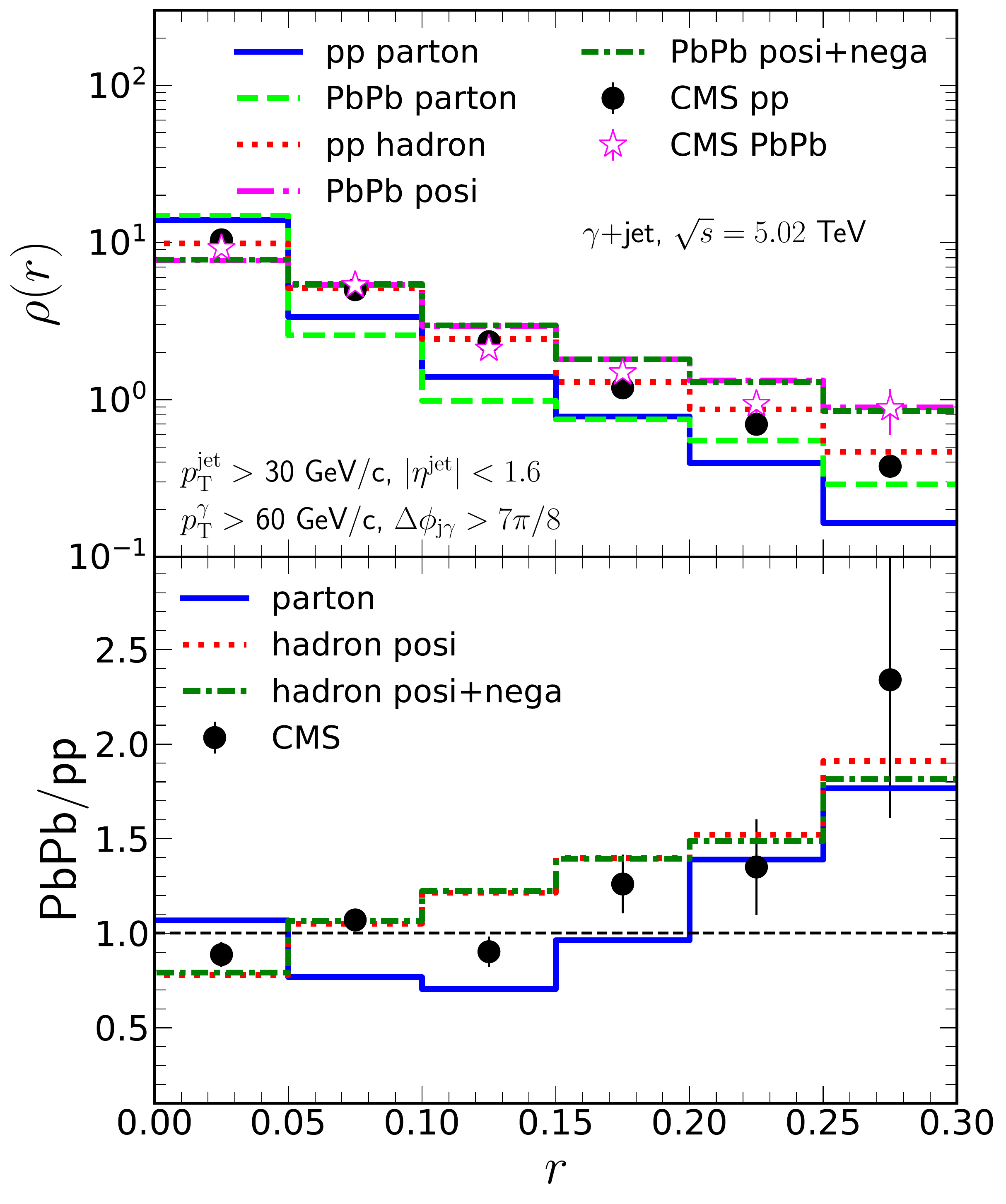}
\vspace{-5pt}
  \caption{(Color online) Top:  jet shapes  $\rho(r)$  for $\gamma$ tagged jets   in  p+p  and Pb+Pb collisions. Bottom: the ratio of  this jet shape in Pb+Pb to that in  p+p  collisions. Experimental data are taken from CMS measurement~\cite{CMS:2018jco}. } \label{jeshape}
\end{figure}

To compare with experimental data~\cite{CMS:2018jco,Sirunyan:2018qec}, we use the same kinematic cut as CMS experiment~\cite{CMS:2018jco,Sirunyan:2018qec}. In the numerical computation, $\gamma$ is required in the region $|\eta^\gamma|<$1.44, and $p_\text{T}^{\gamma}>$60 GeV/$c$. Background contributions from fragmentation and decay
photons are rejected by imposing the isolation requirements: the $p_\text{T}$ sum of final hadrons in a cone of radius 0.4 with respect to the centroid of the photon is required to be less than 1 GeV/$c$. The anti-$k_\text{T}$ jet finding algorithm~\cite{Cacciari:2011ma,Cacciari:2008gp} is used to cluster the resulting particles using a distance parameter R=0.3. Jets with $|\eta^\text{jet}|<$1.6 and $p_\text{T}^\text{jet}>$30 GeV/$c$ are selected. An azimuthal separation
of $\Delta\phi_{\text{j}\gamma} =|\Delta\phi_\text{j}-\Delta\phi_{\gamma}| > 7\pi/8 $ is applied to select the most balanced  photon-jet
pairs. The charged hadrons are collected with requirement~\cite{CMS:2018jco,Sirunyan:2018qec} that $p_\text{T}^\text{ch,h} >$1 GeV/$c$, $|\eta^\text{ch,h}|<$ 2.4, and must fall within a cone of radius $\Delta r=\sqrt{(\phi^\text{ch,h}-\phi^\text{j})^2+(\eta^\text{ch,h}-\eta^\text{j})^2}<$0.3 around the jet direction.

\begin{figure}[t]
\centering
\includegraphics[width=0.45\textwidth]{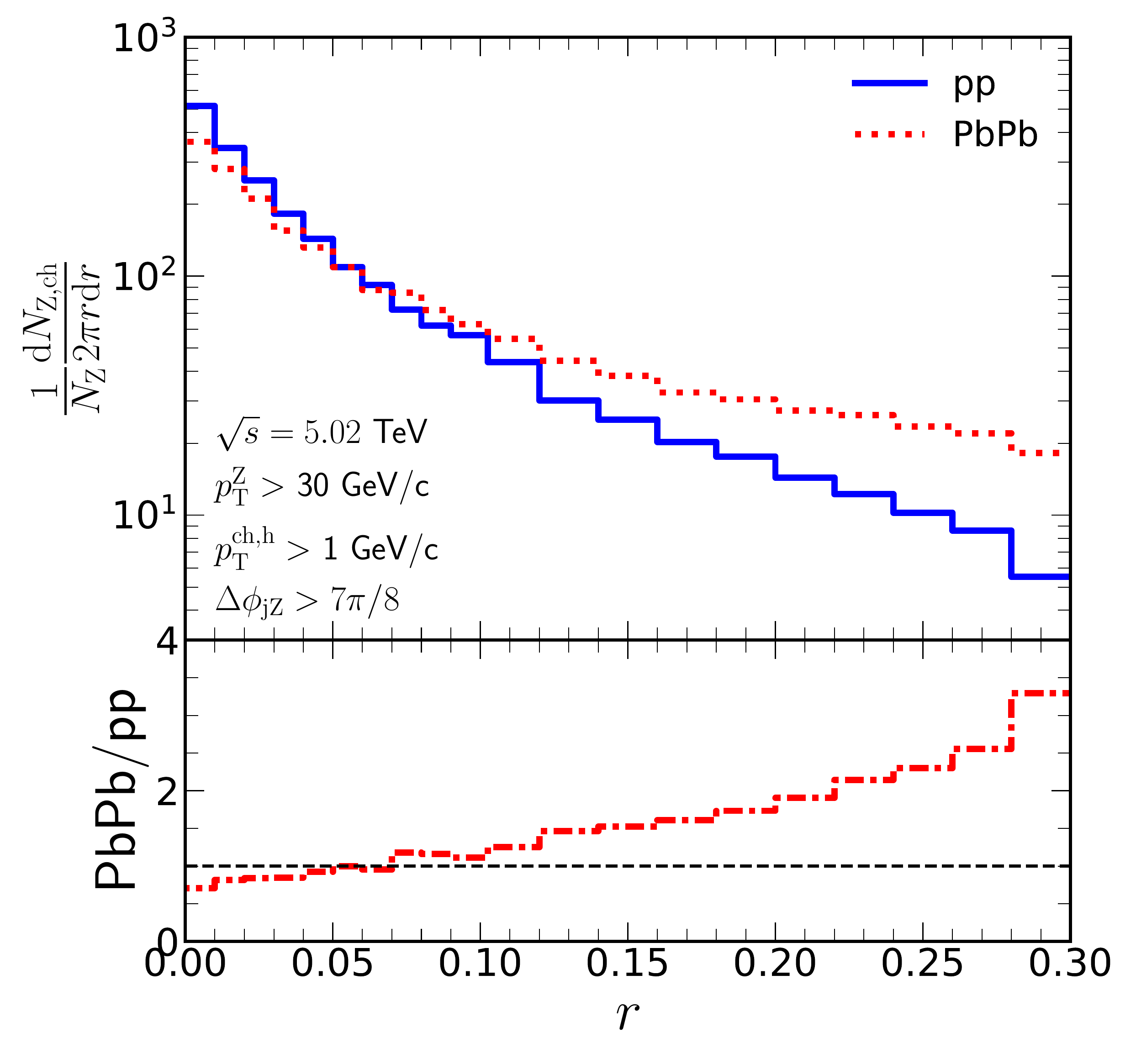} \\
\vspace{-0pt}
  \caption{(Color online) Top: distributions of particle number density of jets associated with a photon as a function of the distance to the jet axis in   p+p   and Pb+Pb collisions. Bottom: the ratio of charged number density in Pb+Pb to that in   p+p  collisions. } \label{number_density}
\end{figure}

To have a detailed understanding of how the energy is redistributed inside and outside the jet in Pb+Pb, we first calculate the differential jet shape for jets associated with an isolated photon as defined in Eq.~\ref{eq:jeshape}.
The top panel in Fig.~\ref{jeshape} shows the differential jet shape of isolated photon tagged jet calculated by our model for both   p+p   and Pb+Pb collisions and the comparison with experimental data~\cite{CMS:2018jco}. The results are normalized to unity over $r<0.3$.  Our model calculations can well describe the jet shapes both  in   p+p   and Pb+Pb collisions.
The bottom panel in Fig.~\ref{jeshape}  presents the ratio of the jet shape in Pb+Pb to that in  p+p  collisions.
A depletion is observed in the region  $r<0.05$ and an enhancement in the region $r > 0.1$ in the central Pb+Pb collisions relative to  p+p.
Jet-medium interactions lead to a larger fraction of jet energy carried by soft particle at large distance from the jet axis, and result in medium modified jet energy distribution in heavy-ion collisions.   We also see that negative partons make rather insignificant contributions to jet shape in Pb+Pb. This is because, jet shape is the  radial average transverse momentum  distributions inside the jet.  In the calculation of jet shape, only charged hadrons with transverse momentum $p_\text{T}>1$ GeV/$c$ is considered and the  hadrons from negative partons that fall into the jet cone are rather few. Furthermore, the momentum of those hadrons from negative partons is relatively small compared to the jet momentum.

To further illustrate the effect of hadronization on the medium modification of jet substructures, we also show  the distributions along with the nuclear modification factors of $\rho(r)$  calculated at parton level in Fig.~\ref{jeshape}. As can be seen, the ratios demonstrate moderately different trend at  parton level compared to hadron level, indicating that the suppression of jet energy density near jet axis is underestimated at parton level. The difference can be explained by two reasons.  
For one thing, as mentioned above, the reference spectrum of $\rho(r)$ is quite different between parton level and hadron level in p+p collisions.
For another, the effect of hadronization on jet substructures in p+p collisions is greater  than that in  A+A collisions, because the distance of the color connected positive partons  will become closer in Pb+Pb collisions due to medium-induced radiation compared to p+p collisions.  Therefore, the effect of hadronization is essential for the medium modification on jet energy redistribution in heavy-ion collisions,  indicating the hadronization effect may play a very important role for the possible explanation  of jet cone dependent $R_\text{AA}$ puzzle observed so far in heavy-ion collisions~\cite{alice:jetcone,star:jetcone,CMS:2021vui,ATLAS:2012tjt}.

In addition to the transverse momentum distribution within and out of jet cone,  we predict the particle number density
  $n_\text{ch}(r)=\frac{1}{N_\text{jet}}\frac{dN_\text{ch}}{2\pi r dr}$,  where $r$ is the distance between changed particles and jet axis in the $\phi-y$ plane. Distributions of $n_\text{ch}(r)$ in   p+p   and Pb+Pb collisions  are shown in  the top panel of Fig.~\ref{number_density}.
  The number density distributions is quite similar  with jet shape. Most of the particles are distributed in the jet core. 
  The ratio of the number density in Pb+Pb collisions to that in p+p collisions goes up smoothly as the increasing of $r$. 
   Compared to  p+p   collisions,  charged hadrons density is suppressed near jet axis, while increased by  two times on the brink of jets due to jet-medium interactions.

\begin{figure}[t]
\centering
\vspace{-10pt}
\includegraphics[width=0.45\textwidth]{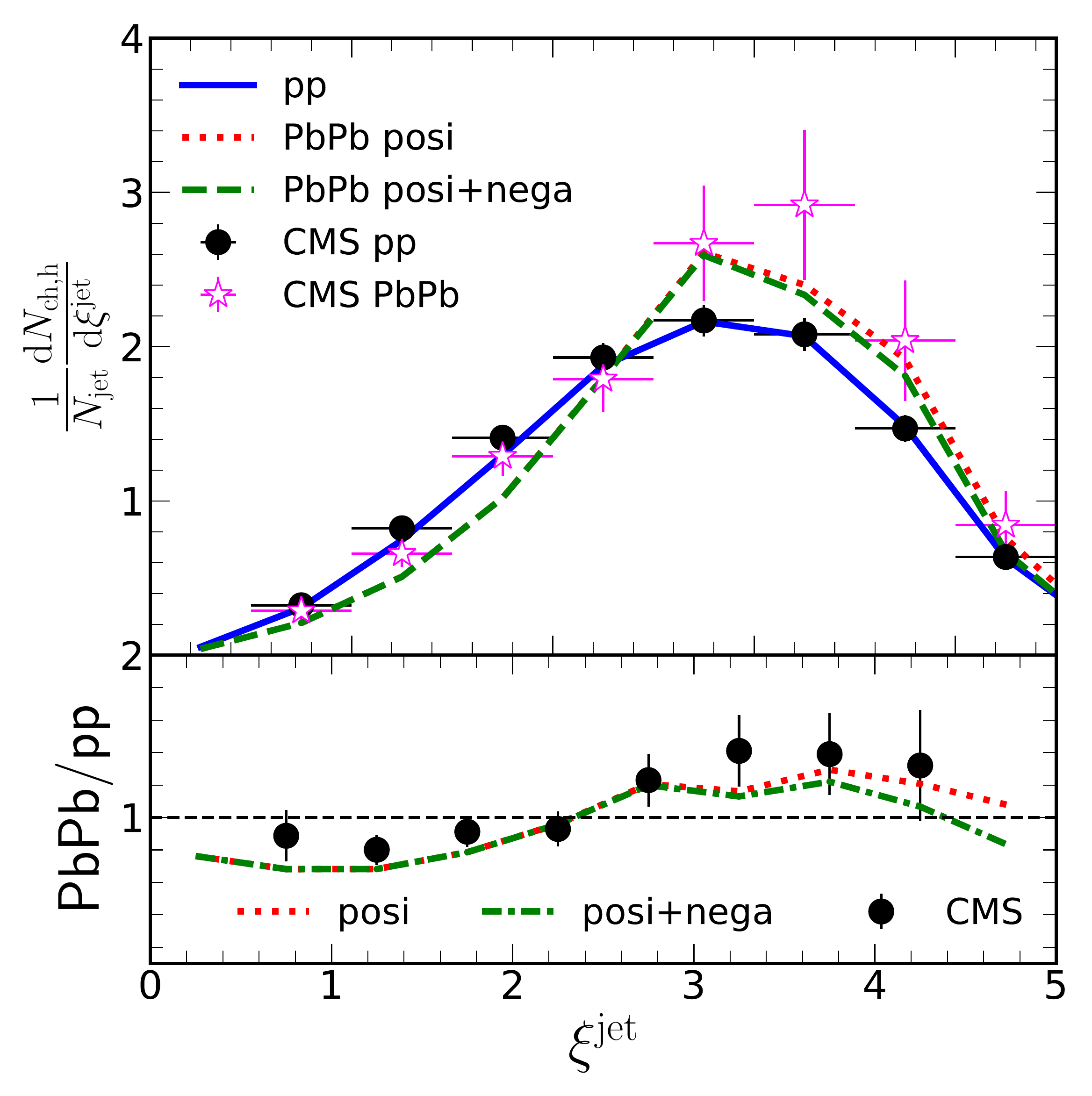} 
\includegraphics[width=0.45\textwidth]{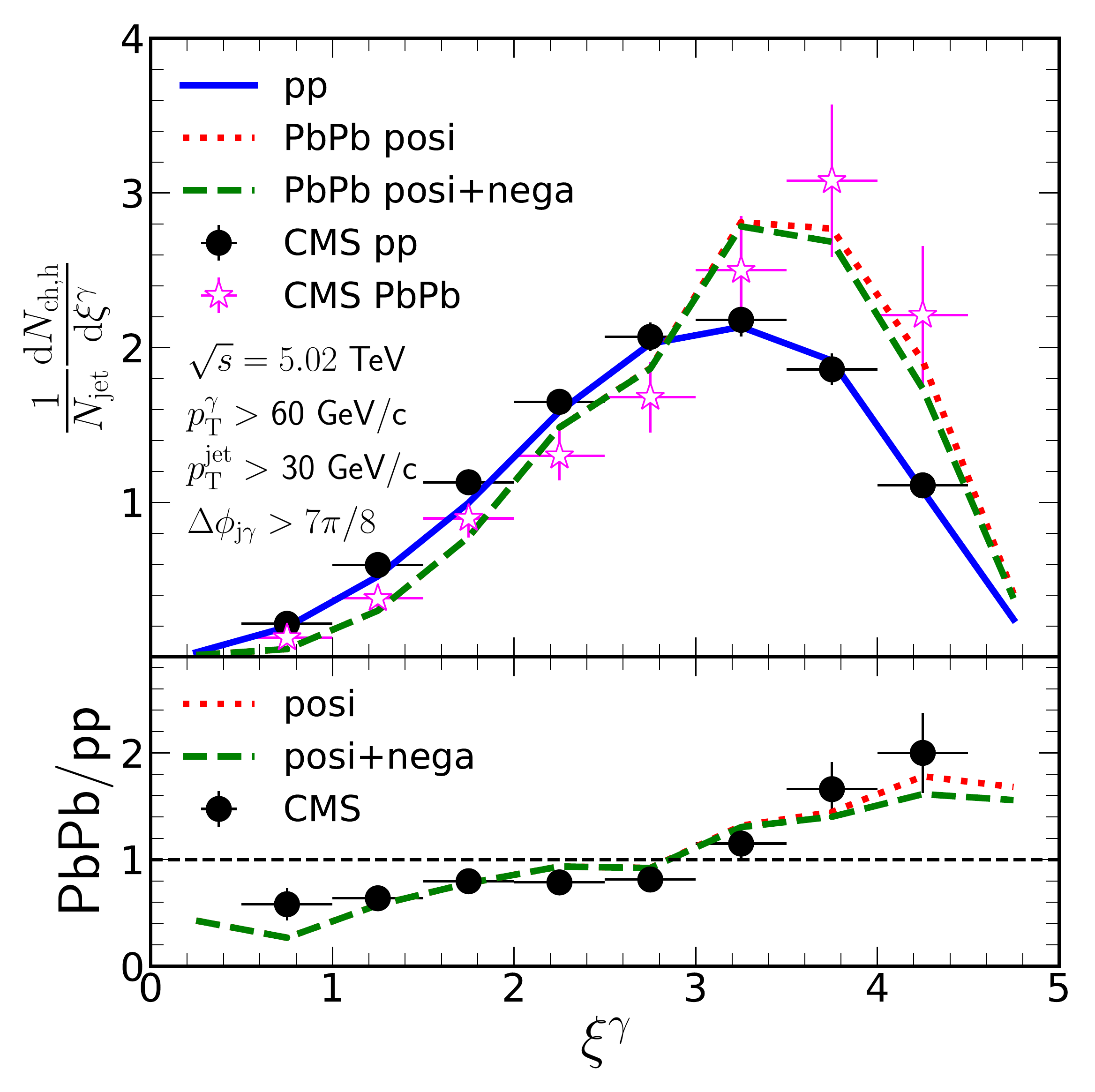} 
\vspace{0pt}
  \caption{(Color online) Top: Distributions of jet fragmentation  function $\xi^\text{jet}$ in  p+p  and Pb+Pb collisions  and the ratio of  $\xi^\text{jet}$ in Pb+Pb  to that in  p+p  collisions. Bottom: distributions of jet fragmentation  function $\xi^{\gamma}$ in  p+p  and Pb+Pb collisions (up panel) and the ratio of  $\xi^{\gamma}$ in Pb+Pb  to that in  p+p  collisions. Experimental data are taken from Ref.~\cite{Sirunyan:2018qec} } \label{pjet_fragmentaion}
\end{figure}

In top panel of Fig.~\ref{pjet_fragmentaion}, we plot the fragmentation function of jet (see the definition in Eq.~\ref{eq:jet_frag}) in association with  $\gamma$ for  p+p  and Pb+Pb collisions.    The ratio of the fragmentation function in Pb+Pb to that in  p+p  collisions is shown in  Fig.~\ref{pjet_fragmentaion}. Our model describes the experimental data~\cite{Sirunyan:2018qec} well both in  p+p  and Pb+Pb collisions.    One can see a moderate suppression in the region $\xi^\text{jet}<$2 and a significant enhancement in region $\xi^\text{jet}>$2. 
 The  definition of $\xi^\text{jet}$ in Eq.~\ref{eq:jet_frag} tells that jet fragmentation $\xi^\text{jet}$  is a kind of the particle number density inside the jet, for example, small $\xi^\text{jet}$  region corresponds to high energy hadrons and large $\xi^\text{jet}$ region comes from soft hadrons. Therefore, the increase of soft hadrons due to jet-medium interactions lead to the  enhancement in region $\xi^\text{jet}>$2 in Pb+Pb collisions.  Similarly as jet  shape,  $\xi^\text{jet}$ is nonsensitive to hadrons from negative partons.

\begin{figure}[t]
\centering
\vspace{-10pt}
\includegraphics[width=0.465\textwidth]{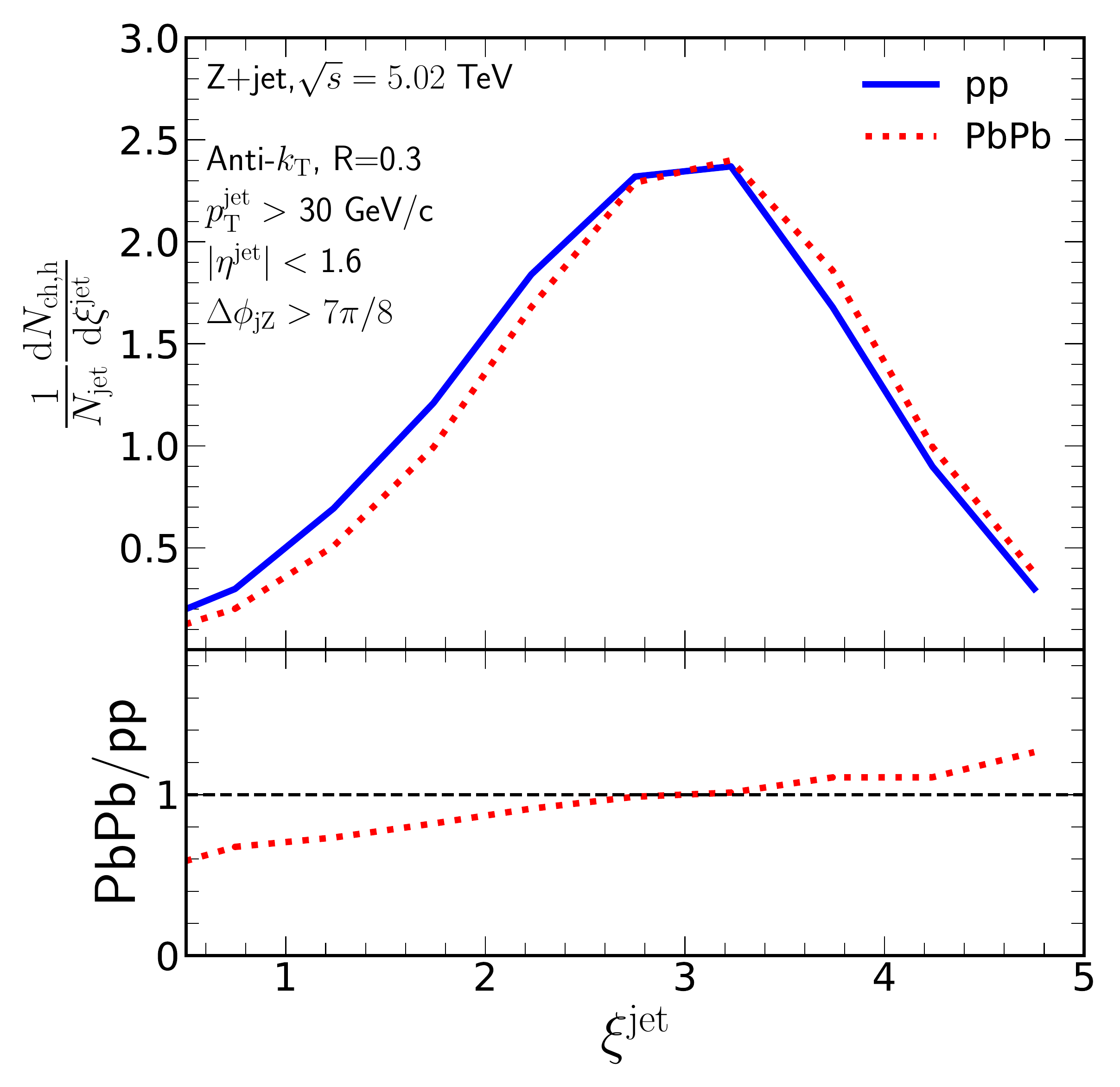}
\includegraphics[width=0.465\textwidth]{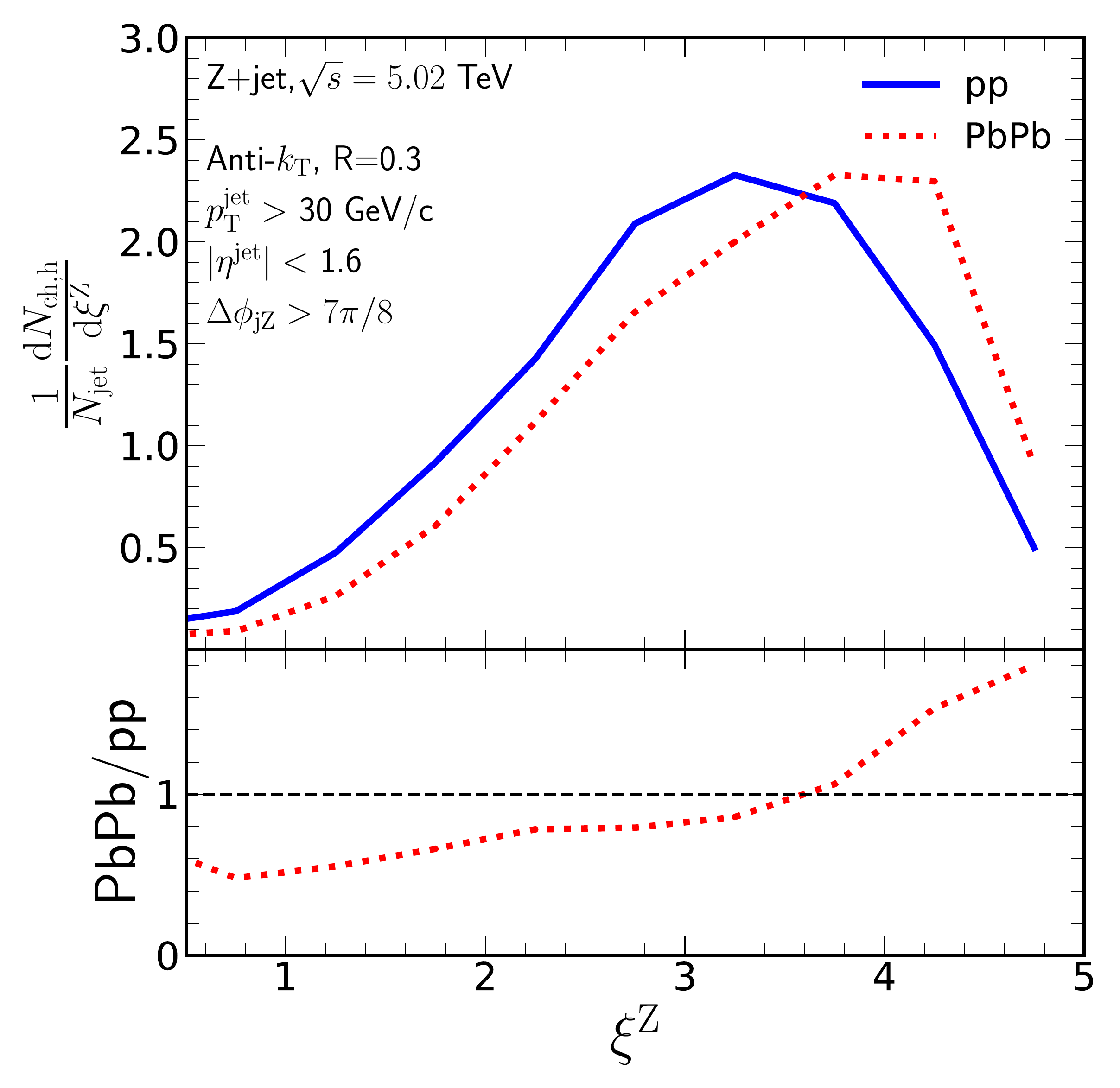} 
 \caption{$Z^0$-jet fragmentation functions respect to jet momentum (up) and $Z^0$ boson transverse momentum (bottom) in  p+p and Pb+Pb collisions  and the ratios in Pb+Pb to that in p+p collisions. } \label{Zjet_frag}
\end{figure}

In  photon tagged jet production, since photon does not participate in strong interaction directly and has a mean free path much longer than the size of the QGP, its energy will be intact during the propagation in the medium. The energy loss of the jet partons will be better characterized with the medium modification of the jet fragmentation function as~\cite{Wang:2013cia,Chen:2020tbl}:
\begin{equation}
\xi^{\gamma}_\text{T}=\ln \frac{-|\overrightarrow{p}^{\gamma}_\text{T}|^2}{\overrightarrow{p}^\text{ch,h}_\text{T}\cdot\overrightarrow{p}^{\gamma}_\text{T}}
\end{equation}

The variable $\xi^{\gamma}_\text{T}$ gives the fragmentation pattern with respect to the transverse momentum of the initial parton. The bottom picture in Fig.~\ref{pjet_fragmentaion} shows the $\gamma$-jet fragmentation functions  $\xi^{\gamma}_\text{T}$  for  p+p  and Pb+Pb collisions as well as their ratio, which  are in a good agreement with the experimental data.   Compared to $\xi^\text{jet}$ in  p+p  collisions, in the Pb+Pb collisions an enhancement is observed  in the $\xi^{\gamma}_\text{T} > 3$ region, while a suppression in the region $\xi^{\gamma}_\text{T} < 3$. The medium modification is much more significant on $\xi^{\gamma}_\text{T}$ distributions than $\xi^\text{jet}$ distributions. This suppression and enhancement of the jet fragmentation pattern are direct evidence of energy loss of high energy partons propagating through the hot-dense medium created in heavy-ion collisions.  Hadron yields from negative partons hadronization give moderate contributions to large $\xi^{\gamma}_\text{T}$ and  $\xi^\text{jet}$ due to their low energies.




Finally, we provide theoretical predictions on $Z^0$ associated jet fragmentations,  and plot in Fig~\ref{Zjet_frag} fragmentation distributions with respect to the momentum of the reconstructed jet  $\xi^\text{jet}$ and  with respect to the transverse momentum of recoiling $Z^0$ boson $\xi^{Z}_\text{T}$ in  p+p  and Pb+Pb collisions.  Relative to  p+p  collisions, charged hadron yields with large transverse momentum are suppressed and soft charged hadron yields are enhanced in Pb+Pb collisions.

\section{Conclusions}
\label{sec:conclusion}

We have made a detailed study of medium modification on $Z^0/\gamma$+hadron correlations  and  $Z^0/\gamma$ tagged jet substructures in Pb+Pb collisions relative to  p+p  collisions at 5.02 TeV. The initial reference jet in p+p collisions is generated by SHERPA with NLO+PS,  jet-medium interactions and medium response in the hot-dense medium are simulated by LBT, and the non-perturbative transition of quarks and gluons into final hadrons are taken into account within an extend cluster hadronization model. Based on this framework, we can well describe the experimental data of $Z^0/\gamma$+hadrons correlations and jet substructures in both  p+p  and Pb+Pb collisions.

 Investigations  on the $\Delta \phi_\text{Z,ch}$ in different $p_\text{T}^\text{ch,h}$ intervals  show that the soft hadron yields are significantly increased in Pb+Pb collisions and sensitive to medium response.  Charged hadrons from negative partons are uniformly distributed in the whole phase space. However, charged hadrons from positive partons are Gaussian distributed with mean value $\Delta \phi_\text{Z,ch}=\pi$.  Distributions for high energy charged hadron  is narrowed and suppressed  in Pb+Pb collisions relative to p+p collisions, whereas distributions for soft hadrons are broadened  and enhanced due to jet-medium interactions. The azimuthal angle correlations $\Delta \phi_\text{Z,ch}$ distribution between charged hadrons and the recoiling $Z^0$   is dominated by soft particles and sensitive to hadrons from negative partons, and the increasing of soft hadron yields lead to the overall enhancement of $\Delta \phi_\text{Z,ch}$ in  Pb+Pb collisions relative to p+p collisions.  
 And  transverse momentum distributions $p_\text{T}^\text{ch}$ of charged hadrons  are enhanced in low $p_\text{T}$ region and suppressed for high $p_\text{T}$ hadrons,  $\xi^\text{Z,ch}_\text{T}=\ln [-|\overrightarrow{p}^{Z}_\text{T}|^2/(\overrightarrow{p}^\text{ch,h}_\text{T}\cdot\overrightarrow{p}^{Z}_\text{T})]$ distributions are shifted to larger value.

Furthermore, the medium-modification on jet energy density $\rho(r)$, jet hadron number density $n_\text{ch}(r)$,  and jet fragmentation functions $\xi^{\gamma}_\text{T}$ and  $\xi^\text{jet}$ are  studied in Pb+Pb collisions. Our numerical results demonstrate an excellent agreement with  experimental data. Medium modification on jet profile and jet fragmentation functions indicate that
particles carrying a large fraction of the jet momentum are generally closely aligned with the jet axis, whereas low-momentum particles are observed to have a much broader angular distribution relative to jet axis in Pb+Pb collisions. For completeness,
jet fragmentation  functions respect to jet momentum and $Z^0$ boson transverse momentum for $Z^0$ boson tagged jet are predicted.

 Meanwhile, we find that hadronization effects are not only indispensable in precise calculations of  the spectrum of some jet substructures, but also are  essential for the medium modification on jet energy redistribution in heavy-ion
collisions.  The difference  of nuclear modification factor between parton level and hadron level show that  the hadronization effects also reduce parton energy, and compete with jet quenching, indicating that  the effects of hadronization  are also important for the possible solution of jet cone dependent $R_\text{AA}$ puzzle as seen in measurements of Pb+Pb collisions at the LHC~\cite{alice:jetcone,star:jetcone,CMS:2021vui,ATLAS:2012tjt}.

Our work provides a valuable attempt to generalize the cluster hadronization mechanism to study jet productions in heavy-ion collisions and pave a new way to studying the  non-perturbative hadronization processes with the existence of the QGP. With some simple assumptions of  hadronization of positive partons and negative partons in medium, our results can well describe many observables of the charged hadrons and jet substructures in both p+p and Pb+Pb collisions.  It will be of great interest to investigate different types of hadron yields, such as $\pi$, $p$, $ K $, $\eta$ ~\cite{Ma:2018swx,Zhang:2022fau} and heavy  flavor hadrons~\cite{Wang:2021jgm} in heavy-ion collisions with the extended cluster model, which can be utilized to further constrain and improve the hadronization model presented here.

\vspace{1.0cm}
\section{Acknowledgements}
This research is supported by Guangdong Major Project of Basic and Applied Basic Research No. 2020B0301030008,
by National Natural Science Foundation of China (NSFC) under Grants  No. 12147131, No. 11935007, No. 12035007 and No. 12022512. S.Z. is also supported by the MOE Key Laboratory of Quark and Lepton Physics (CCNU) under Project
No. QLPL2021P01.

\end{document}